\documentclass[letterpaper,10pt, conference]{IEEEtran}

\usepackage[utf8]{inputenc}
\usepackage{graphicx}
\usepackage{caption}

\usepackage{cite}
\usepackage{subcaption}
\usepackage{sidecap}
\usepackage{amsthm,amsmath}
\usepackage{framed}
\usepackage[font={footnotesize,it}]{caption}
\usepackage{color}
\usepackage{marvosym}
\usepackage{amssymb}
 
\theoremstyle{definition} 

\newcommand{\TT}{\,\hat{T}^u}
\newcommand{\TTd}{\,\hat{T}^d}
\newcommand{\DS}{\,\displaystyle }

\newcommand{\deff}{\,\stackrel{\text{def}}{=}}

\newcommand{\eq}[1]{\,\begin{equation}
                   #1 
                   \end{equation}
}

\newcommand{\ave}[1]{\,\displaystyle \left \langle #1 \right \rangle}
\newcommand{\fracc}[2]{\,\displaystyle \frac{#1}{#2}}

\newcommand{\morabba}[1]{\,\begin{flushright}
 \Rectsteel \\
\end{flushright}}

\newcommand{\eqq}[2]{\,\begin{equation} \label{#2}
                   #1 
                   \end{equation}
}

\usepackage{setspace}
\newcommand{\al}[1]{\,\begin{align}
                   #1 
                   \end{align}
}
\newcommand{\all}[2]{\,\begin{align}
                   #1 
                    \label{#2}
                   \end{align}
}
\newcommand{\rond}[2]{\, \displaystyle \frac{\partial #1}{\partial #2}}
\newcommand{\rondd}[2]{\, \displaystyle \frac{\partial^2 #1}{\partial #2^2}}

\usepackage[top=24mm, bottom=25mm, left=22mm, right=20mm]{geometry}

\begin{document}

\title{Voter Model with Arbitrary Degree Dependence:
  Clout, Confidence and Irreversibility}

\author{Babak Fotouhi and Michael G. Rabbat \\
Department of Electrical and Computer Engineering\\
McGill University, Montr\'eal, Qu\'ebec, Canada\\
Email:\texttt{ babak.fotouhi@mail.mcgill.ca, michael.rabbat@mcgill.ca}\\}

\maketitle

\begin{abstract}
In this paper, we consider the voter model with popularity bias. The influence of each node on its neighbors depends on its degree. We find the consensus probabilities and expected  consensus times for each of the states. We also find the fixation  probability, which is the probability that a single node whose state differs from every other node imposes its state on the entire system. In addition, we find the expected fixation time. Then two extensions to the model are proposed and the motivations behind them are discussed. The first one  is confidence, where in addition to the states of neighbors, nodes take their own state into account at each update. We  repeat the calculations for the augmented model and investigate the effects of adding confidence to the model. 
The second proposed extension is irreversibility, where one of the states is given the property that once nodes adopt it, they cannot switch back. The dynamics of densities, fixation times and consensus times are obtained.  

\end{abstract}

\section{Introduction}\label{sec:intro}
To study social systems, one can focus on collective large-scale  social phenomena (macro  behavior) or on the behavior of individuals and their interactions (micro behavior). Macro social behaviors are simultaneously consequences and determinants of micro behaviors. For example, culture is a product of many individual actions, and in turn affects the action of each individual. The recent upsurge in network science has led to a panoply of models of opinion dynamics. These models are agent-based and seek to unify the micro-macro duality~\cite{CS1,CS2,CS3,CS4,CS6,CS7}. Contributions and applications also  exist in diverse strands of research such as  
economics~\cite{econ1,econ5},  biology~\cite{bio1,bio2}   and physics~\cite{phy1,phy2,phy3}.

In models of opinion dynamics, each node has a `state'. This state can be  continuous~\cite{olfati,jad1,acem_cont_1, cont_new}, or discrete~\cite{liggett1,ozdag_heter,yildiz_stub}. One of the most studied models is the voter model, where states can take only two possible values. 
The voter model was  first introduced in~\cite{liggett1} (hereafter, we call the version of the voter model in~\cite{liggett1}  the `pure' voter model). It is a simple stochastic process which models dynamics of dissensus and consensus between agents (hereafter, nodes). One  rationale behind this model is the fact that people are evidently influenced by their peers when they make decisions~\cite{asch}, and that their observations and   interactions with others can affect their behavior remarkably~\cite{sherif}.

In the voter model, nodes are endowed with dichotomous states, typically denoted by ${\pm 1}$. This simplified representation is applicable to situations where there are two choices to take, e.g., seeing or not seeing a film,  choosing between  two major political stances, and whether or not adopt a new product or technology.  Time increments are discrete. At each timestep, one node is randomly selected to update its state. A fraction of the neighbors of this node adopt state~$+1$ and another fraction adopt state~$-1$ at that instant. These fractions constitute two probabilities, and the node adopts one of the states with respective probabilities. 


The voter model has been studied on lattices  in different dimensions~\cite{redner_book} and arbitrary heterogeneous networks~\cite{SOOD, ozdag_heter}. 
Typical lines of inquiry include the probability that the system will reach eventual unanimity on either of the states, and the expected time it takes to reach those states. This model has been also generalized to include three states~\cite{three_voter}. In~\cite{yildiz_stub, zealot_1, zealot_2}, the existence of stubborn nodes are considered. These nodes never alter their states. In~\cite{inertia_voter}, nodes have inertia. The longer a node stays with the same state, the less likely it gets for that node to  change it. In~\cite{noise}, time-dependent transition rates are considered through a noise-reduction scheme. In~\cite{convict_1,convict_2}, nodes have heterogeneous conviction (or persuasion). In~\cite{adapt_1,adapt_2,adapt_3} an `adaptive' network is envisaged, where links whose incident nodes are of opposing states are rewired so that the new link has two agreeing nodes at its ends. In~\cite{GOOD}, a popularity bias is incorporated into the model. At each timestep, a link is selected and then with probabilities that depend on the degrees of incident nodes to that link, one of them imposes its state on the other one. The information that is considered to be known  about the topology of the  underlying network varies among models. For example, in~\cite{ozdag_heter, yildiz_stub, CS6}  the adjacency matrix is assumed to be known completely, whereas in~\cite{SOOD,GOOD,MF} only some statistics of the network are known, such as the degree distribution, in addition to the assumption that the network is connected (see~\cite{MF} for a thorough discussion).

In this paper, we consider that each node assigns `weights' to the state of its neighbors (equivalent to~\cite{GOOD,VW}). If a neighbor has degree ~$k$, the weight its state will have is denoted by $f(k)$. This extension could be used to emulate the  idea that   in social settings, the nodes with higher degrees are more `central' and cast more influence on the decision making of their peers. We find the probability of consensus on either state, expected time to reach unanimity, and the expected time to reach consensus conditional on the final state. Also, we find the `fixation probability', which is the probability that   a system  in which all nodes are unanimous except one anti-conformist node will eventually reach consensus on the state adopted by that single node. 

The pure voter model lacks an important element  present in realistic social settings, which is confidence. In this paper, we endow the  nodes with confidence. This means that each node, in addition to the state of its neighbors, accounts for  its own (weighted) state as well. This extension is proposed to remedy a peculiarity of the  voter model. Consider a star graph, where the central node has state $+1$ and degree $k$. Let all the $k$ peripheral nodes have state $-1$. In the pure voter model, the probabilities of reaching consensus on either $+1$  or  $-1$ for this system  are  both equal to $\frac{1}{2}$~\cite{SOOD}. This is at odds with the intuitive  expectation from social interactions.  In social networks, \emph{opinion leaders} are typically characterized by such centrality. Such central nodes usually have `clout', and influence others' opinions heavily. In the degree-dependent voter model without confidence, this problem persists, because all neighbors of the central node have the same  degree, and weighting them as a function of degree  will give identical weights. However, as shall be described, if we endow nodes with confidence, so that they account for their own opinions as well, this problem is resolved and the central node will have a higher chance of imposing its state, compared to the peripheral nodes. 

Another scenario that the pure voter model is not applicable to is evoked in marketing applications. Let nodes with state $+1$ represent those who have seen a film, and nodes with state $-1$ represent those who have not. In this case, the voter model must be modified, since a node with state $+1$ cannot `un-see' a film and flip back to $-1$. To emulate this setting, we propose an `irreversible' version of the voter model, where nodes with state $+1$ cannot flip back to $-1$ (note that the case with no degree-dependence is akin to the SI model of disease epidemics~\cite{bio2}). We initiate the system with a certain fraction of nodes adopting state$-1$ at the outset (which in realistic cases is typically small, because most of the population have not seen the film), and study the dynamics of the system. We find the time required for the system to reach unanimity, and the time it takes for a single node with state $+1$ to impose it on the entire network. 
%
%
%


\section{The Model}
Let the nodes be connected on the graph ${G=(V,E)}$. We do not assume that  $E$ is known. We assume that~$G$ is connected and that the degree distribution is known.  
Let us denote the state of node ${x \in V}$ at time $t$ by $s_x(t)$. States are binary, so that $s_x(t)= \pm 1$. States are updated asynchronously. 
We  take the time unit to be equal to $\frac{1}{N}$, where $N$ is the number of nodes. Consequently, at each timestep, on average all nodes are selected once to update their states.
At each timestep, node $x$ calculates probabilities to adopt each of the two states at the next timestep. Denote the probability that node $x$ will adopt $+1$ and $-1$ at time ${t+1}$ by $P_x^+(t+1)$ and $P_x^-(t+1)$, respectively. Let $\mathcal{N}_x $ denote the set of neighbors of node $x$. Also let $\mathcal{N}_x^+(t)$ and $\mathcal{N}_x^-(t)$ denote the set of neighbors of $x$ with state $+1$ and $-1$ respectively. For any node $y$, let $z_y$ denote its degree, i.e., the number of its neighbors. The update scheme is modeled as follows:
\all{
\begin{cases}
P_x^+(t+1)= \displaystyle \frac{\sum_{y \in \mathcal{N}_x^+} f(z_y)}{\sum_{y \in \mathcal{N}_x}  f(z_y)} \\ \\
P_x^-(t+1)= \displaystyle \frac{\sum_{y \in \mathcal{N}_x^-} f(z_y)}{\sum_{y \in \mathcal{N}_x}  f(z_y)}.
\end{cases}
}{Pha}
Note that if $f(\cdot)$ is a constant, then these probabilities become the fraction of nodes adopting each of the states, which is the case for the pure voter model~\cite{liggett1}. 

We  assume that the underlying graph is connected. We analyse the model within the framework used in~\cite{SOOD,GOOD}, which is observed to perform remarkably well  for scale-free graphs with negligible degree correlations~\cite{MF,SOOD,VW}. 

 Throughout, the total number of nodes is denoted by $N$. 
We denote the fraction of all nodes that are adopting the state $+1$ at time~$t$ by $\rho(t)$. Number of nodes with degree $k$ is denoted by $N_k$, and the fraction of nodes with degree $k$ is denoted by $n_k$. So $n_k$ denotes the degree distribution of the graph.  The fraction of  nodes with degree $k$ whose state at time $t$ is $+1$ is denoted by $\rho_k(t)$. 

%

Denote the expected value of the state of node $x$ at time $t+1$ by $c_x(t+1)$. 
This quantity can be obtained by the weighted average of the neighbors 
of node~$x$ in the following way
\eqq{
c_x(t+1)= \displaystyle \frac{\sum_{y \in \mathcal{N}_x} c_y(t) f(z_y)}{\sum_{y \in \mathcal{N}_x} f(z_y)}
.}{cx_1}

Let $\langle \phi(k) \rangle$ denote the network average of the quantity $\phi(k)$, that is, ${\sum_k n_k \phi(k)}$. For example, $\langle k \rangle$ denotes the average degree of the graph. 
Now we find the expected value of the numerator and the denominator 
 of~\eqref{cx_1}, assuming that the degree-degree correlations of the underlying graph is negligible, which is true for random graphs realized, for example, in~\cite{SFU1,SFU2,SFU3}, and works for a broad range of heterogeneous networks~\cite{MF}. For the denominator, we have
 \all{
 \ave{\sum_{y \in \mathcal{N}_x}  f(z_y)}
 = z_x \sum_k \fracc{k n_k}{\ave{k}} f(k) = \fracc{z_x}{\ave{k}} \ave{k f(k)}
 . 
 }{denom_1}
For the numerator, we have
 \all{
 \ave{\sum_{y \in \mathcal{N}_x} c_y(t) f(z_y)}
 = z_x \sum_k \fracc{k n_k}{\ave{k}} f(k) \big[ 2\rho_k(t)-1\big] 
 . 
 }{num_1}
Let us define the following weighted density 
\all{
\mu(t) \deff \displaystyle \sum_k k n_k f(k) \rho_k (t)
.}{mu_def}
So~\eqref{num_1} can be rewritten as follows
 \all{
 \ave{\sum_{y \in \mathcal{N}_x} c_y(t) f(z_y)}
 = \fracc{z_x}{\ave{k}} \bigg[2 \mu(t) - \ave{k f(k)} \bigg] 
 . 
 }{num_2}
 
Using this together with~\eqref{denom_1}, we can express~\eqref{cx_1}   equivalently as follows
\all{
\ave{k f(k)} c_x(t+1)&= 2 \displaystyle    \sum_k  k n_k  f(k)  \rho_k (t) 
- \ave{k f(k)}
\nonumber \\
&= 2 \mu(t) - \ave{k f(k)}
.}{cx_2}
Let us multiply both sides by $z_x f(z_x)$ and then sum over all $x$.  On the right hand side we will have the sum~${\sum_x z_x f(z_x)}$, which is equal to~$N \ave{k f(k)}$. On the left hand side we will have  the sum~${\sum_x z_x f(z_x) c_x(t+1)}$. This sum can be expressed equivalently as follows
\all{
\displaystyle \sum_{x \in V} z_x f(z_x) c_x(t+1) &= 
N \displaystyle \sum_k k n_k f(k) \big[2 \rho_k(t+1)-1  \big]  
\nonumber \\
&= N \bigg[2 \mu(t+1) - \ave{k f(k)}\bigg] 
.}{sigma_mu}
So after multiplying by the factor~$z_x f(z_x)$ and summing over all~$x$, Equation~\eqref{cx_2} transforms into the following
\al{
& \ave{k f(k)}  N \bigg[2 \mu(t+1) - \ave{k f(k)}\bigg]
\nonumber \\
& = 
N  \bigg[2 \mu(t) - \ave{k f(k)} \bigg]  \ave{k f(k)}
.}
This implies that ${\mu(t+1)=\mu(t)}$, so that the quantity $\mu(t)$ is conserved.

\subsection{Consensus Probability} \label{sec:prob_uncorr}

The conservation of $\mu(t)$ leads us to the probability of eventual consensus on each of the states. Let us define the consensus probabilities
\eq{
\begin{cases}
P^u \stackrel{\text{def}}{=} P\{\rho(\infty)=1\}  \\
P^d \stackrel{\text{def}}{=}  P\{\rho(\infty)=0\}
\end{cases}
.}

\begin{figure}
  \centering
  \includegraphics[width=1\columnwidth]{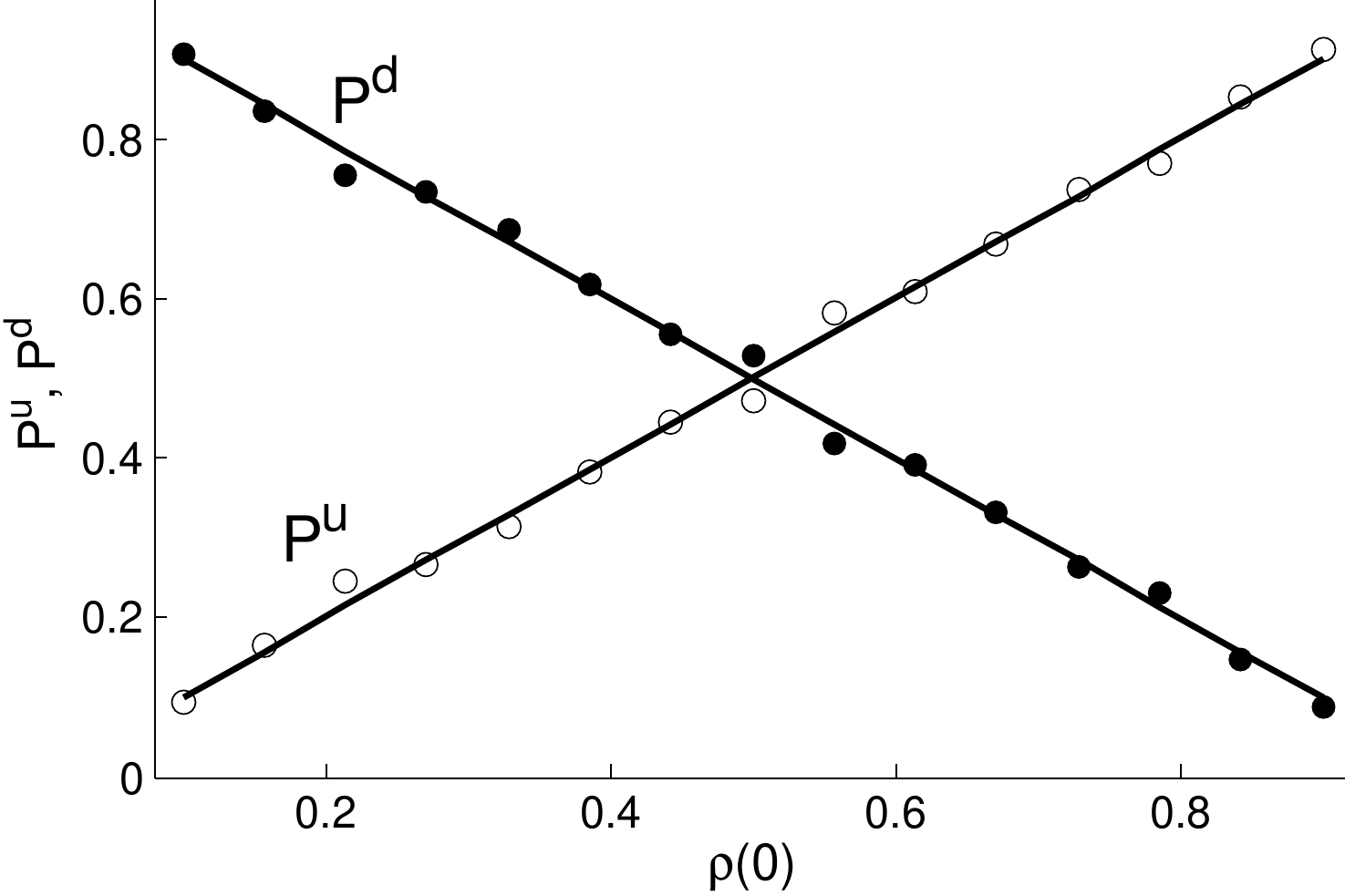}
  \caption[Figure ]%
   {  Consensus probability for states $+1$ and $-1$ as a function of initial density of nodes with state $+1$. The markers are simulation results and the solid lines are theoretical predictions given by~\eqref{pu_2}  and~\eqref{pd_2}. The underlying graph has power law degree distribution ${n_k \sim k^{-3}}$ as proposed in~\cite{BA}, with $m=20$. The special case of ${f(k)=k^{0.8}}$ is considered. The network has 1500 nodes and the results are averaged over 100 Monte Carlo trials.}
\label{ordinary_P}
\end{figure}

At~${t \rightarrow \infty}$ the value of $\mu$ is the same as that at~${t=0}$. When the system is at state~${\rho=1}$, then the value of $\mu$ is equal to $\ave{k f(k)}$. When $\rho$ is zero, then the value of $\mu(t)$ is equal to zero. From conservation of $\mu(t)$,  we have:
\al{
\mu(0)=\big[ \ave{k f(k)} \big] P^u  + [ 0  ] P^d  
.}
For the probability of eventual consensus on state $+1$ we obtain
\all{
P^u = \fracc{\mu(0)}{ \ave{k f(k)}}= \fracc{\displaystyle \sum_k k n_k f(k) \rho_k(0)}{\displaystyle \sum_k k n_k f(k) }
.}{pu_1}
As~\eqref{sigma_mu} indicates, we can express $\mu(0)$  in terms of the initial conditions as follows
\al{
\mu(0) = \fracc{\displaystyle \sum_{x  \in V}z_x f(z_x) s_x(0)+N\displaystyle \sum_{x  \in V}z_x f(z_x) }{2N}
.}
Plugging this expression into~\eqref{pu_1}, we find the following alternative form of the eventual consensus probability 
\all{
P^u= \frac{1}{2} + \fracc{\displaystyle \sum_{x  \in V}z_x f(z_x) s_x(0)}{\displaystyle \sum_{x  \in V}z_x f(z_x) } 
.}{pu_2}
Similarly, the eventual consensus probability on state $-1$ is obtained
\all{
P^d= \frac{1}{2} - \fracc{\displaystyle \sum_{x  \in V}z_x f(z_x) s_x(0)}{\displaystyle \sum_{x  \in V}z_x f(z_x) } 
.}{pd_2}
In Fig.~\ref{ordinary_P}, theoretical predictions are compared to simulation results. The underlying graph is one obtained by the preferential attachment scheme proposed in~\cite{BA}, which yields graphs with power-law degree distributions with exponent $3$. We used $m=20$, which means that in the sequential construction of the graph, each new node attaches to $20$ existing nodes with a mechanism described in~\cite{BA}. A network of size ${N=1500}$ is used for simulations. The special case of ${f(k)=k^{0.8}}$ is considered as an example.

Note that in the case of $f(\cdot)$ being a constant, which is synonymous with the conventional voter model~\cite{liggett1,SOOD}, for $\mu$ we have $\mu= \sum_k k n_k \rho_k$, and the two consensus probabilities are simplified to~${P^u= \fracc{\mu(0)}{\ave{k}}}$~ and~${P^d= 1-\fracc{\mu(0)}{\ave{k}}}$. 
These agree with the results for the conventional voter model, obtained in~\cite{SOOD}.

\subsection{Time to Consensus} \label{sec:time_uncorr}

In this section we will first focus on the expected time to reach unanimity. We denote this time by~$T$. In the next section we will turn our attention to the expected time to reach consensus on state $+1$ and that to reach state $-1$. We denote these two quantities by $T^u$ and $T^d$, respectively. 

We begin by introducing the  increment and decrement probabilities for nodes of degree $k$. The increment probability is the probability that a node of degree $k$ flips its state from $-1$ to $+1$, and the decrement probability is defined vice versa.  When these events occur, the fraction of nodes of degree $k$ with state $+1$ changes by ${\delta \rho_k=\frac{1}{N_k}}$.  We define
\eq{
\begin{cases}
P^+_k  \stackrel{\text{def}}{=} P\{ \rho_k \rightarrow \rho_k + \delta \rho_k \}  \\
P^-_k \stackrel{\text{def}}{=}   P\{ \rho_k \rightarrow \rho_k - \delta \rho_k \}  
\end{cases}
.}

Let us represent the state of the system by a vector $\vec{\rho}$, where 
$\vec{\rho} (t)= (\rho_1(t), \rho_2(t), \rho_3(t), \ldots)$ encodes the densities of sub-populations of nodes with different degrees. Let $\mathbf{e}_k$ denote the unit vector along the $k$-th dimension. 
We denote the expected time to reach unanimity by $T(\vec{\rho})$.
 If a node of degree $k$ flips its state, then $\rho_k$ will vary by $\frac{1}{N n_k}$. Let us denote this change by $\delta \rho_k$. 
 
The following recurrence relation holds for the expected unanimity time: 
\all{
T(\vec{\rho})&= \delta t + \sum_k \left[ T(\vec{\rho}+\delta \rho_k \mathbf{e}_k) P^+_k +  T(\vec{\rho}-\delta \rho_k \mathbf{e}_k) P^-_k \right]
\nonumber \\
&\quad
+ \left[ 1- \sum_k  (P^+_k +  P^-_k) \right] T(\vec{\rho})
.}{T_recursive_1}

This equation relates $T$ for densities $\vec{\rho}$ at two successive time steps; the time-dependence is omitted in the expression to improve readability. The first sum on the right hand side accounts for the case that an increment or decrement occurs. The second sum refers to the case where no change occurs. 

Now let us consider the Taylor expansion of~\eqref{T_recursive_1} up to second order. 
After some algebraic steps, we obtain
\all{
0 &= \delta t 
 + \displaystyle \sum_k  \left( P_k^+ - P_k^- \right)     \rond{T(\vec{\rho})}{\rho_k} \delta \rho_k   
\nonumber \\
&\quad + 
\displaystyle \sum_k    \left( P_k^+  +    P_k^- \right)     \rondd{T(\vec{\rho})}{\rho_k} \fracc{\delta \rho_k^2}{2}  
.}{T_PDE_2}

We can simplify this equation further. 
Using the chain rule  
 and rearranging the terms, we arrive at the following equation
\all{
& \rond{T(\mu)}{\mu} \displaystyle \sum_k  \left( P_k^+ - P_k^- \right)      k   f(k)     
\nonumber \\
&+ 
\rondd{T(\mu)}{\mu}  \displaystyle \sum_k    \left( P_k^+  +    P_k^- \right)     k^2   [f(k)]^2 \fracc{1}{2N}  =-1
.}{T_PDE_4}

To continue, we need the increment and decrement probabilities. For the increment probability, we have
\all{
P^+_k = n_k (1-\rho_k) \fracc{\displaystyle \sum_{\ell} \frac{\ell n_{\ell}}{\ave{k}} f(\ell) \rho_{\ell}}{\displaystyle \sum_{\ell} \frac{\ell n_{\ell}}{\ave{k}} f(\ell) }
 = n_k (1-\rho_k) \fracc{\displaystyle \mu}{\ave{k f(k)} }
.
}{P_inc}
The factor $n_k (1-\rho_k)$ is the portion of all nodes who have degree $k$  and state $-1$. The factor that multiplies this fraction is the probability  that  a node whose state is $-1$ switches to state $+1$. 
Similarly, for the decrement probability we have
\all{
P^-_k = n_k  \rho_k \fracc{\displaystyle \sum_{\ell} \frac{\ell n_{\ell}}{\ave{k}} f(\ell) (1-\rho_{\ell})}{\displaystyle \sum_{\ell} \frac{\ell n_{\ell}}{\ave{k}} f(\ell) }
 = n_k  \rho_k \bigg[ 1- \fracc{\displaystyle \mu}{\ave{k f(k)} } \bigg] .
}{P_dec}
Using~\eqref{P_inc} and~\eqref{P_dec}, 
%
 we can simplify \eqref{T_PDE_4}. We get
\all{
&  \rond{T(\mu)}{\mu} \displaystyle \sum_k  \left[ \fracc{\displaystyle \mu}{\ave{k f(k)} } - \rho_k \right] 
     k n_k  f(k)     
\nonumber \\
&+ 
 \rondd{T(\mu)}{\mu} \displaystyle \sum_k  \Bigg\{  \fracc{k^2 n_k  [f(k)]^2}{2N}  \nonumber \\
 & \times  
 \left[ \fracc{\displaystyle \mu}{\ave{k f(k)} } + \rho_k - 2 \rho_k \fracc{\displaystyle \mu}{\ave{k f(k)} }  \right] \Bigg\}
 =-1
.}{T_PDE_5}
For the first term on the left hand side, note that the following is true
\al{
&\displaystyle \sum_k  \left[ \fracc{\displaystyle \mu}{\ave{k f(k)} } - \rho_k \right] 
     k n_k  f(k)     
\nonumber \\
&= 
\fracc{\displaystyle \mu}{\ave{k f(k)} } \displaystyle \sum_k  k n_k  f(k)  
- \displaystyle \sum_k  k n_k  f(k)  = \mu-\mu=0
.}
So~\eqref{T_PDE_5} simplifies to 
\all{
\!\!\!
 \rondd{T(\mu)}{\mu} \displaystyle \sum_k \!\!    \fracc{k^2 n_k  [f(k)]^2}{2N}  
\left[  \fracc{\displaystyle -\mu}{\ave{k f(k)} } - \rho_k + \fracc{\displaystyle  2 \rho_k \mu}{\ave{k f(k)} }   \right] 
 =1
.}{T_PDE_6}

To simplify this equation further, note that the expected change in $\rho_k$ is ${P^+_k-P^-_k}$. So we have
\al{
\frac{d}{dt} \rho_k =   \fracc{\displaystyle \mu}{\ave{k f(k)} } - \rho_k 
.}
Integrating this equation yields
\al{
\rho_k(t) - \fracc{\displaystyle \mu}{\ave{k f(k)} }  =  
 \left[ \fracc{\displaystyle \mu}{\ave{k f(k)} } - \rho_k (0)\right]
 e^{-  t}
.}
This means that the deviation of $\rho_k$ from ${\fracc{\displaystyle \mu}{\ave{k f(k)} }}$ decays exponentially fast in time. This happens for all values of $k$. 
After this rapid change, the dynamics of the system enters a more slowly-varying phase~\cite{SOOD,GOOD}.

Using this fact, we can approximate~\eqref{T_PDE_6} by confining the range of time   to the second phase of the dynamics. 

 So the differential equation for the unanimity time~\eqref{T_PDE_6} transforms into the following
\all{
   \rondd{T(\mu)}{\mu}  
\fracc{\displaystyle \mu}{\ave{k f(k)} }  \left[ 1-   \fracc{\displaystyle \mu}{\ave{k f(k)} }  \right]  
 =\fracc{-N}{\ave{k^2  f(k) ^2}}
.}{T_PDE_8}
Let us define the new variables
\all{
\begin{cases}
\xi \deff \fracc{\displaystyle \mu}{\ave{k f(k)} }
\\ \\
\Lambda \deff \fracc{N \ave{k  f(k)}^2}{\ave{k^2  f(k) ^2}}.
\end{cases}
}{lambda_defs}
In terms of the new variables,~\eqref{T_PDE_8} transforms into the following  
\all{
   \rondd{T(\xi)}{\xi}  
\xi  \left( 1-   \xi  \right)  
 =- \Lambda
.}{T_PDE_9}
In order to solve this second-order differential equation we require two boundary conditions. Note that when all nodes are at state $+1$, i.e.,  when $\xi$ is equal to unity, then $T$ will be zero. Also when all nodes have state $-1$, which means that $\xi$ is zero, then $T$ will be equal to zero. Thus the two boundary conditions are~${T(1)=T(0)=0}$. 
Integrating~\eqref{T_PDE_9} twice and applying these boundary conditions, we obtain

%
\all{
  T(\xi)    
 =- \Lambda \bigg[\xi \ln \xi   +(1-\xi)   \ln(1-\xi)  \bigg] 
.}{T_FIN}
Replacing  $\Lambda$ by its explicit expression given in~\eqref{lambda_defs}, we obtain
\all{
  T(\xi)    
 = \fracc{-N \ave{k  f(k)}^2}{\ave{k^2  f(k) ^2}} \bigg[\xi \ln \xi   +(1-\xi)   \ln(1-\xi)  \bigg] 
.}{T_FIN_full}

\begin{figure}
  \centering
  \includegraphics[width=1\columnwidth]{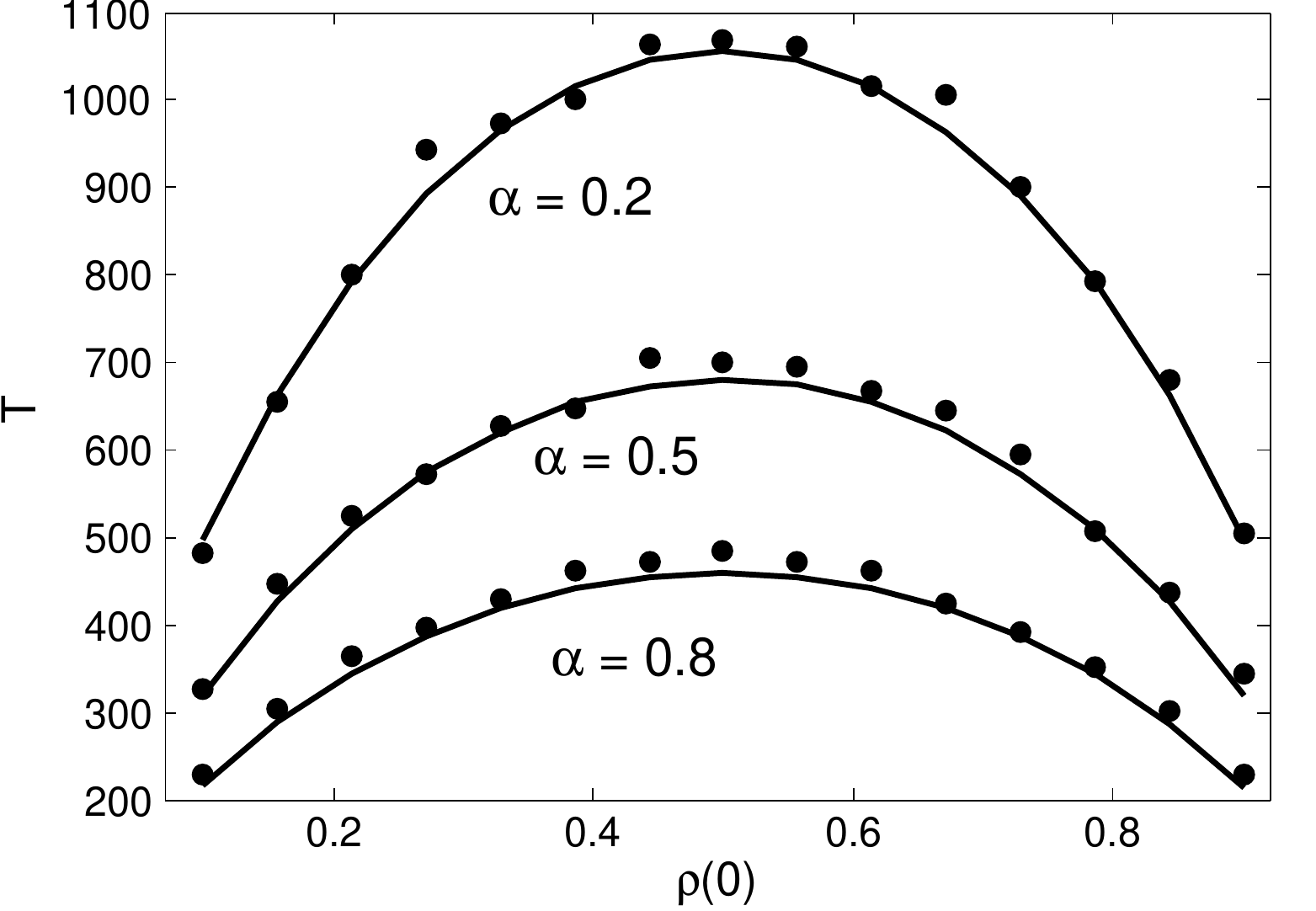}
  \caption[Figure ]%
   { Expected time to reach unanimity as a function of initial density of nodes with state $+1$. The markers are simulation results and the solid lines are theoretical predictions given by~\eqref{T_FIN_full}. The underlying graph has power law degree distribution ${n_k \sim k^{-3}}$ is proposed in~\cite{BA} with $m=20$. The network has 1500 nodes and the results are averaged over 100 Monte Carlo trials. The case of ${f(k)=k^{\alpha}}$ is considered with ${\alpha=0.2,0.5,0.8}$. As can be seen from the graph, higher values of $\alpha$ reach unanimity faster. This means that assigning more influence to nodes of larger degrees expedites convergence.
   }
\label{ordinary_T}
\end{figure}

This agrees with the result obtained in~\cite{GOOD,VW} for link update dynamics. Theoretical prediction and simulation results are compared in Fig.~\ref{ordinary_T}. The underlying network is a  scale-free graph~\cite{BA} with $m=20$. The number of nodes is $1500$ and the results are compared for ${f(k)=k^{\alpha}}$ with ${\alpha=0.2,0.5,0.8}$.

\subsection{Time to Reach Consensus Conditional on the Final State}
First let us focus on the expected time to reach consensus on the state $+1$, which we denote by $T^u$. Since this time is conditional upon  the eventual consensus being on state $+1$, we have to make adjustments to~\eqref{T_recursive_1}. The recurrence equation becomes:
\all{
P^u(\vec{\rho}) T^u(\vec{\rho})&= P^u(\vec{\rho}) \delta t 
\nonumber \\
&+ \sum_k \Bigg[ P^u(\vec{\rho}+\delta \rho_k \mathbf{e}_k) T^u(\vec{\rho}+\delta \rho_k \mathbf{e}_k) P^+_k
\nonumber \\
&  + P^u(\vec{\rho}-\delta \rho_k \mathbf{e}_k) T^u(\vec{\rho}-\delta \rho_k \mathbf{e}_k) P^-_k \Bigg]
\nonumber \\
&
+ \left[ 1- \sum_k  (P^+_k +  P^-_k) \right] P^u(\vec{\rho}) T^u(\vec{\rho})
.}{Tu_recursive_1}

Let us define
\eqq{
\begin{cases}
\TT (\vec{\rho}) \deff  P^u(\vec{\rho}) T^u(\vec{\rho})
\\ \\
\TTd (\vec{\rho}) \deff  P^d(\vec{\rho}) T^d(\vec{\rho})
\end{cases}
.}{TT_def}
Then~\eqref{Tu_recursive_1} reduces to the following equation
\all{
0&= P^u(\vec{\rho}) \delta t 
\nonumber \\
&+ \sum_k \left(P^+_k-P^-_k  \right) \rond{\TT(\mu)}{\mu} k n_k f(k) \delta \rho_k
\nonumber \\
&
+  \sum_k \left(P^+_k + P^-_k  \right) 
\rondd{\TT(\mu)}{\mu} k^2 n_k^2 f(k)^2 \fracc{\delta \rho_k^2}{2}
.}{Tu_recursive_2}

We proceed by restricting the time domain within the second phase of the dynamics similar to the previous stage. Thus~\eqref{Tu_recursive_2} simplifies to the following
\all{
- P^u(\vec{\rho}) \delta t &= \sum_k \Bigg[\left(1-  \fracc{\mu}{\ave{k f(k)}} \right) \nonumber \\
&\times
\rondd{\TT(\mu)}{\mu} k^2 n_k^3 f(k)^2 \delta \rho_k^2 \fracc{\mu}{\ave{k f(k)}}
\Bigg]
.}{Tu_recursive_3}
Writing this equation in terms of~$\xi$ and~$\Lambda$ defined in~\eqref{lambda_defs}, and using~\eqref{pu_1} for the term on the left hand side, and multiplying both sides by $N$, this equation reduces to
\all{
\rondd{\TT(\xi)}{\xi} = -\fracc{\Lambda}{1-\xi}
.}{Tu_recursive_4}
Integrating this equation twice and dividing by~${P^U}$, we obtain
\all{
T(\xi) = \frac{-\Lambda}{\xi} \Bigg[(1-\xi) \ln(1-\xi)-(1-\xi)+K_3 \xi + K_4 \Bigg] 
,}{Tu_recursive_5} 
where $K_3$ and $K_4$ are integration constants. 
 The first boundary condition is~$T(1)=0$, which gives
${K_3=-K_4}$. So we obtain
\all{
T^u(\xi) = -\Lambda \fracc{(1-\xi)}{\xi} \Bigg[ \ln(1-\xi)-1-K_3  \Bigg] 
.}{Tu_result_2}
Note that symmetry of the dynamics readily determines the following result 
\all{
T^d(\xi) = -\Lambda \fracc{ \xi}{1-\xi} \Bigg[ \ln( \xi)-1-K_3  \Bigg] 
.}{Td_result_2} 

To determine $K_3$ note that the following holds
\eq{
P^u T^u+P^d T^d=T.
}
Plugging in the results obtained in~\eqref{Tu_result_2} and~\eqref{Td_result_2} into the left hand side and using the expression for $T$ in~\eqref{T_FIN} on the right hand side, 
we obtain 
$K_3=-1$. After replacing the explicit expression for $\Lambda$ given in~\eqref{lambda_defs},  we arrive at the conditional consensus times 
\all{
\begin{cases}
T^u(\xi)=  \fracc{-N \ave{k  f(k)}^2}{\ave{k^2  f(k) ^2}} \fracc{1-\xi}{\xi} \ln(1-\xi) \\ \\
T^d(\xi)=\fracc{-N \ave{k  f(k)}^2}{\ave{k^2  f(k) ^2}}  \fracc{\xi}{1-\xi} \ln(\xi) 
\end{cases} 
}{T_COND_FIN}

\begin{figure}
  \centering
           \begin{subfigure}[b]{1\columnwidth}
       \centering    
  \includegraphics[width=1\columnwidth]{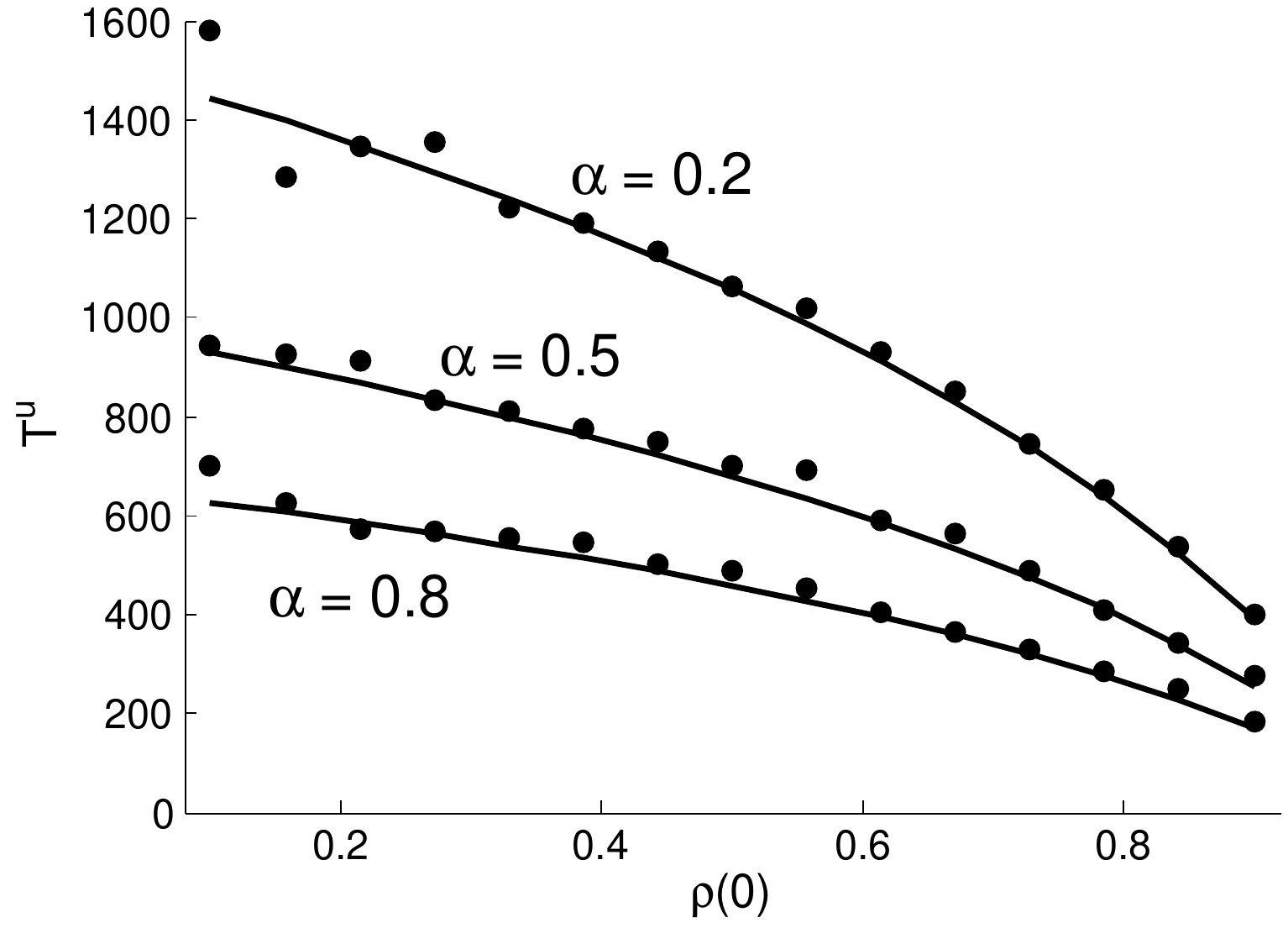}
  \caption{ }
                \label{fig:T_up}
   \end{subfigure}
   
    \begin{subfigure}[b]{1\columnwidth}
       \centering    
  \includegraphics[width=1\columnwidth]{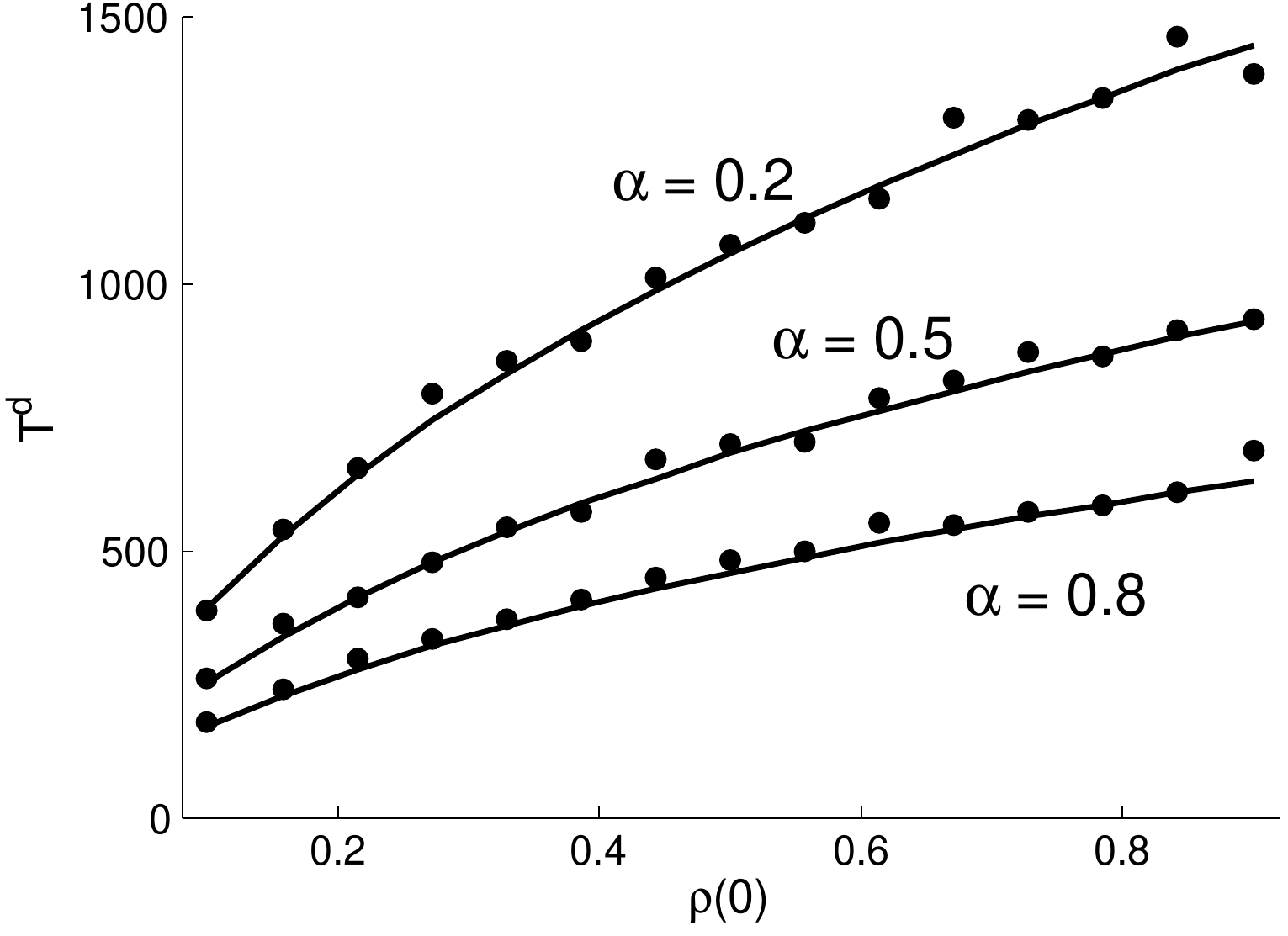}
  \caption{ }
                \label{fig:T_down}
   \end{subfigure}
   
  \caption[Figure ]%
   { Expected time to reach unanimity conditional on consensus over (a) state~$+1$, and (b) state~$-1$. Conditional consensus times are depicted as a function of initial density of nodes with state $+1$. The markers are simulation results and the solid lines are theoretical predictions given by~\eqref{T_COND_FIN}. The underlying graph has power law degree distribution ${n_k \sim k^{-3}}$ as proposed in~\cite{BA} with $m=20$. The network has 1500 nodes and the results are averaged over 100 Monte Carlo trials. The case of ${f(k)=k^{\alpha}}$ is considered with ${\alpha=0.2,0.5,0.8}$. As can be seen from the graph, higher values of $\alpha$ reach unanimity faster. This means that assigning more influence to nodes of larger degrees expedites convergence. This was also observed in Fig.~\ref{ordinary_T}.
   }
\label{ordinary_TU}
\end{figure}

Theoretical predictions are presented along with simulation results in Fig.~\ref{ordinary_TU}.

%
%

 \subsection{Example: Linear Clout} \label{sec:pref_comp}
 Now let us focus on a simple case of~$f(k)$ that can encode popularity bias, namely $f(k)=k$. Popularity bias means that when node~$x$ is updating its state, the higher the degree of a neighbor is, the more influence that neighbor will have on the state to be adopted by node~$x$. Linear popularity bias is referred to as \emph{linear clout} hereafter
 
As an example, let us investigate the dependence of the expected time to reach unanimity for the special case of graphs with power-law degree distribution whose degree-degree correlations are negligible (through the recipe articulated in~\cite{SFU1,SFU2,SFU3}, for example). For these networks, we have
 \eq{
 n_k \sim k^{-\gamma}
 .}
From~\eqref{T_FIN_full} we see that the quantities $\ave{k f(k)}$ and  $\ave{k^2  f(k) ^2}$ are required to study the behavior of expected unanimity time. For a network with power-law degree distribution, first we estimate the maximum degree, denoted by~$k_{\max}$,  through a heuristic method used in~\cite{SOOD,fluc,SFU3}. For ${\gamma \geq 2}$ we have:
\eq{
\displaystyle \int_{k_{\max}}^{\infty} k^{-\gamma} dk = \frac{1}{N}
.}
This gives~${k_{\max} \sim \DS N^{\frac{1}{\gamma-1}}}$. We focus on the scale-free family of graphs introduced in~\cite{BA} where $\gamma=3$. 
In this case, we have~${\ave{k f(k)} \sim \ln \DS k_{\max} }$. 
So for the numerator of~\eqref{T_FIN_full} we have~${\ave{k f(k)}^2 \sim [ \ln N ] ^2}$.
%
Similarly, for the denominator of~\eqref{T_FIN_full} we obtain~${\ave{k^2 f(k)^2}  \sim N }$.
Inserting these limits into~\eqref{T_FIN_full}, we arrive at
  
\begin{figure}
  \centering
  \includegraphics[width=1\columnwidth]{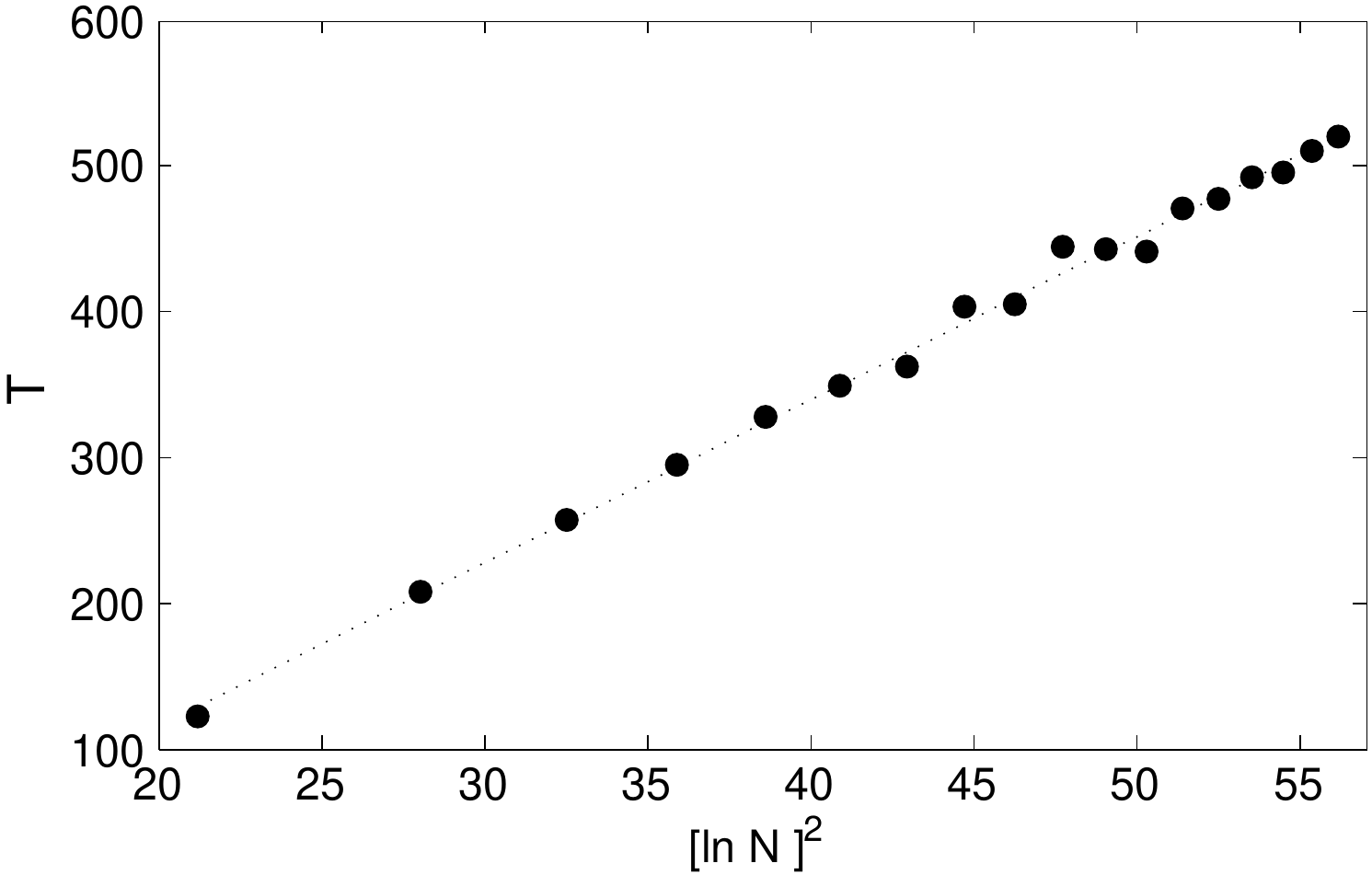}
  \caption[Figure ]%
   { Expected time to reach consensus for the case of linear clout ${f(k)=k}$, depicted as a function of $(\ln N)^2$ as predicted in~\eqref{T_behave_1}. The simulation results are averaged over 1000 Monte Carlo trials.}
\label{alpha_1_T_in_N}
\end{figure}

 \eqq{
T \sim
[\ln N]^2  	,		
}{T_behave_1}

so unanimity is reached faster than the pure voter model~\cite{SOOD}. The unanimity time for the pure voter model is~${T \sim \frac{N}{\ln N}}$\cite{SOOD}, while in the linear clout scheme it is~${T \sim [\ln N]^2 }$. Fig.~\ref{alpha_1_T_in_N} illustrates the  consensus time as a function of $(\ln N)^2$ for  the scale-free networks introduced in~\cite{BA}.

\subsection{Fixation Probability}
Now let us focus on the \emph{fixation probability}, namely the probability of the system reaching consensus on state~$+1$, for the initial condition of a single node at state~$+1$ and all other nodes at state~$-1$. In social contexts, the fixation probability quantifies the likelihood of the emergence of a leader   or the takeover of a minority. In linguistics, it quantifies the probability that a singular way of pronouncing a word overspreads the population~\cite{lingu}. 

We refer to the single deviant node as \emph{ the  mutant}. Suppose that the degree of this node is $z$. This means that~${\rho_z(0)= \frac{1}{N n_z}}$. Let $P^f(z)$ denote the fixation probability for a mutant with degree $z$. 
 Then~\eqref{pu_1} gives the probability of eventual consensus over $+1$ as follows 
\all{
P^f(z) = \fracc{z n_z f(z) \frac{1}{N n_z}}{\displaystyle \sum_k k n_k f(k) } 
= \fracc{1}{N} \fracc{z   f(z)}{\displaystyle \sum_k k n_k f(k) }
.}{p_mut_1}

To proceed, we  consider the case of linear clout over the  uncorrelated networks with power-law degree distribution~${n_k \sim k^{-\gamma}}$. We focus on the case networks grown by preferential attachment as described in~\cite{BA}, for which we have ${\gamma=3}$. For the denominator of~\eqref{p_mut_1}, similar to Section~\ref{sec:pref_comp}, we have~${\ave{k f(k)} \sim   \ln N }$. 
 Assume that the mutant has a small degree, namely ${z \sim O(1)}$. 
Then for the fixation probability we obtain.
\eqq{
\DS P^f(z)  \sim 
\fracc{1}{N  \ln N}   
}{Pf_ordinary}

\begin{figure}
  \centering
  \includegraphics[width=1\columnwidth]{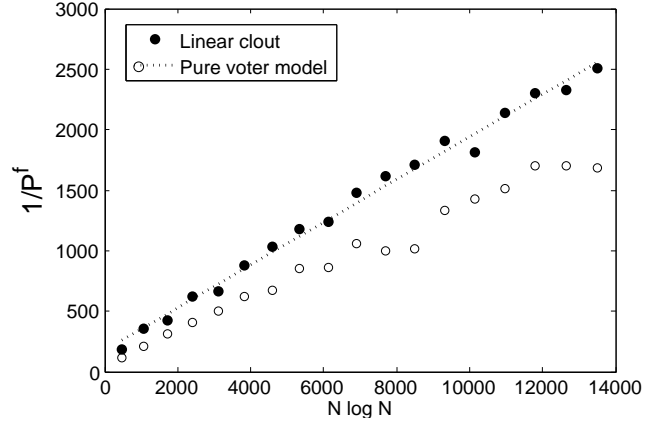}
  \caption[Figure ]%
   {  Depiction of inverse of fixation probability as a function of ${N \ln N}$, which is very close to linear as predicted by~\eqref{Pf_ordinary}. The bottom curve is the depiction of fixation probability for the pure voter model.  
   This figure illustrates the fact that endowing nodes with degree-dependent popularities, reduces the chance of a random single mutant to take over  (note that  inverse of probabilities are depicted, so the bottom one is larger).

The underlying network used in the simulations is scale-free, as introduced in~\cite{BA}, with ${m=20}$. The probabilities are calculated from an ensemble of $10^5$ Monte Carlo trials.
   }
\label{P_behave_NlogN}
\end{figure}

This probability indicates that for large system sizes, the chance for the system to fixate on the state imposed by a single mutant are scant. However, as can be seen from Fig.~\ref{P_behave_NlogN}, this probability is less than that of the pure voter model, for which ${P^f(z) \sim \frac{1}{N}}$.

\section{The Role of Confidence}
As discussed in the introduction, in the pure voter model and also the generalization we considered above, nodes do not account for their own state when they are updating their states. One drawback that stems from this characteristic is the failure to incorporate the model with  confidence. In social settings, nodes have different levels of confidence and persuasiveness, which gives rise to role structures such as  hierarchy of influence~\cite{hier1,hier2}. To be able to model those phenomena, the model of opinion dynamics must incorporate confidence. Here we discuss a generalization to the voter model in order to address this issue.

As an example, consider a star graph where the central node has degree $k$. The central node has state $+1$ and those on the periphery have state $-1$. Suppose nodes do not have confidence, so that each node updates its state according to~\eqref{Pha}. Consider the case of ${f(k)=k^2}$ for expository brevity. In~\cite{SOOD} it is contended that the system reaches consensus over $+1$ or $-1$ with equal probability $\frac{1}{2}$. Now let us add confidence to nodes, which means that they also account for their own states. Now if the central node is chosen to update its state, then the probability of not changing state is $\frac{k^2-k}{k^2+k}$, higher than that of switching, whereas the peripheral nodes adopt the state of the central node with certainty if they are chosen to update state. This elevates the chance of the central node to impose its state over the system above $\frac{1}{2}$, which is closer to the intuitive expectation  that central nodes  have higher influence.

We augment the model by assuming that in the update process, each agent also accounts for its own state. In the pure voter model, this would not change the dynamics significantly, but if there is degree-dependent weighting, then nodes with large degrees will assign a large weight to their own state, which models hierarchy in social influence. 

We begin by modifying~\eqref{cx_1} to account for self-influence as follows:

\eqq{
c_x(t+1)= \displaystyle \frac{  f(z_x) c_x(t) + \DS \sum_{y \in \mathcal{N}_x} c_y(t) f(z_y)}{   f(z_x)+ \DS \sum_{y \in \mathcal{N}_x} f(z_y)}
.}{cx_selfish_1}
%

Inserting~\eqref{num_2} in the numerator and~\eqref{denom_1} in the denominator, we obtain

\eqq{
c_x(t+1)= \displaystyle \frac{  f(z_x) c_x(t)+ \fracc{z_x}{\ave{k}} \bigg[2 \mu(t) - \ave{k f(k)} \bigg] }{ f(z_x)+ \fracc{z_x}{\ave{k}} \ave{k f(k)}}
.}{cx_selfish_2}

  Multiplying both sides by the denominator of the right hand side, and then multiplying both sides by $f(z_x)$, we get
  
\all{
&   c_x(t+1) f^2(z_x) + c_x(t+1) z_x f(z_x) \frac{\ave{kf(k)}}{\ave{k}}
\nonumber \\
&=   \displaystyle   f^2(z_x) c_x(t)+ \fracc{z_x f(z_x)}{\ave{k}} \bigg[2 \mu(t) - \ave{k f(k)} \bigg]  
.}{cx_selfish_3}
Now define (the definition for $\mu$ is repeated here for convenience of reference)
\all{
\begin{cases}
\nu(t) \deff \DS  \sum_k n_k  f^2(k) \rho_k(t) \\
\mu(t) \deff \DS \sum_k n_k k f(k) \rho_k(t) \\
\psi(t) \deff     \nu(t)+ \fracc{\ave{k f(k)}}{\ave{k}} \mu(t).
\end{cases}
}{nu_def}
Now let us sum up~\eqref{cx_selfish_3} over all~$x$ and then divide both sides of the equation by $N$. We get
\all{
& 2    \nu(t+1)-   \ave{f^2(k)} + \frac{\ave{kf(k)}}{\ave{k}} \bigg[ 2 \mu(t+1) - \ave{k f(k)} \bigg]  
\nonumber \\
&= 2   \nu(t)-    \ave{f^2(k)}+ 2 \frac{\mu(t)}{\ave{k}} \ave{k f(k) } - 
\frac{\ave{k f(k) }}{\ave{k}} \ave{k f(k)}
.}{cx_selfish_4}

This can be simplified to
\al{
\psi(t+1)= \psi(t).
}
So the quantity $\psi(t)$ is conserved throughout the dynamics. This immediately leads us to the consensus probability over each of the states.

\subsection{Consensus Probability}
 Note that when all nodes have state $+1$, then~$\psi$ takes the value~${     \ave{f^2(k)} + \frac{\ave{kf(k)}^2}{\ave{k}}}$. Using the conservation of $\psi(t)$, we find that the consensus probability over state $+1$ satisfies
\al{
\psi(0)=   \bigg[   \ave{f^2(k)} + \fracc{\ave{kf(k)}^2}{\ave{k}} \bigg]P^u + 0 P^d
.}
So we obtain
\all{
P^u = \fracc{\psi(0)}{  \ave{f^2(k)} + \frac{\ave{kf(k)}^2}{\ave{k}}}
.}{PU_selfish}
We can use the explicit form for the numerator and obtain the following expression for the consensus probability as a function of initial conditions
\all{
P^u = \fracc{  \DS \sum_k n_k  f^2(k) \rho_k(0) + \fracc{\ave{k f(k)}}{\ave{k}}  \sum_k n_k  k f(k) \rho_k(0)}
{  \DS \ave{f^2(k)} + \frac{\ave{kf(k)}^2}{\ave{k}}}
.}{PU_selfish_full}

Instead of sum over all degrees, we can express this result as a sum over all individual nodes. The result is
\all{
P^u &= \fracc{1}{2}
\nonumber \\
&+ \fracc{  \DS \sum_x z_x  f^2(z_x) s_x(0) + \fracc{\ave{k f(k)}}{\ave{k}}  \sum_x z_x   f(z_x) s_x(0)}
{  \DS  \sum_x z_x  f^2(z_x)   + \fracc{\ave{k f(k)}}{\ave{k}}  \sum_x z_x   f(z_x)  }
.}{PU_selfish_full_2}
Similarly, for consensus over $-1$ we have
\all{
P^d &= \fracc{1}{2}
\nonumber \\
&- \fracc{  \DS \sum_x z_x  f^2(z_x) s_x(0) + \fracc{\ave{k f(k)}}{\ave{k}}  \sum_x z_x   f(z_x) s_x(0)}
{  \DS  \sum_x z_x  f^2(z_x)   + \fracc{\ave{k f(k)}}{\ave{k}}  \sum_x z_x   f(z_x)  }
.}{PD_selfish_full_2}

\begin{figure}
  \centering
  \includegraphics[width=1\columnwidth]{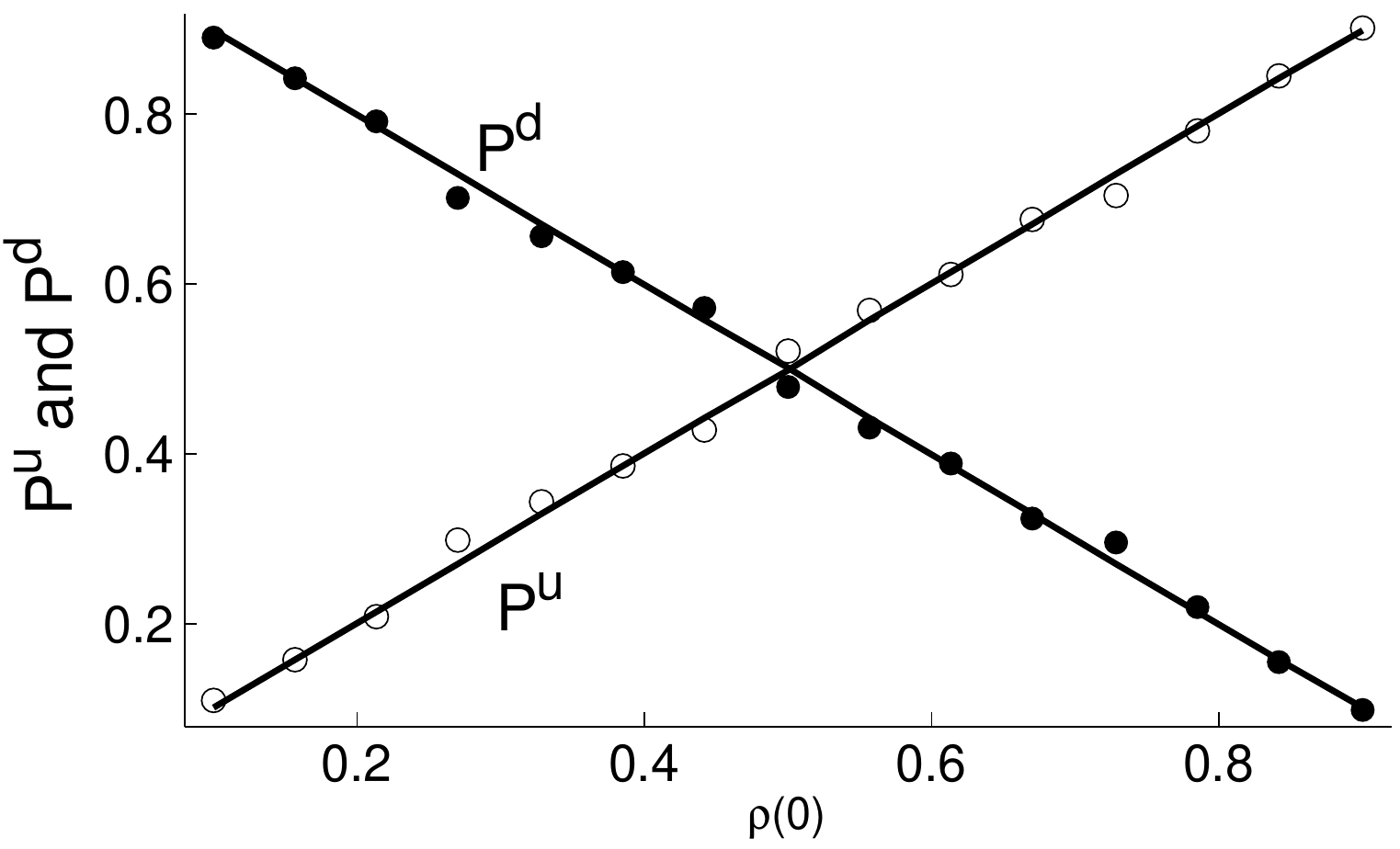}
  \caption[Figure ]%
   {  Consensus probability for states $+1$ and $-1$ as a function of initial density of nodes with state $+1$. The markers are simulation results and the solid lines are theoretical predictions given by~\eqref{PU_selfish_full_2}  and~\eqref{PD_selfish_full_2}. The underlying graph has power law degree distribution ${n_k \sim k^{-3}}$ as proposed in~\cite{BA} with $m=20$. The special case of ${f(k)=k^{0.8}}$ is considered. The network has 1500 nodes and the results are averaged over 100 Monte Carlo trials.}
\label{confident_P}
\end{figure}

Theoretical predictions of~\eqref{PU_selfish_full_2} and~\eqref{PD_selfish_full_2} for consensus probabilities are presented along with simulation results in Fig.~\eqref{confident_P}.

\subsection{Time to Consensus}

In this section, we will take the similar steps undertaken in Section~\ref{sec:time_uncorr} to obtain the expected time to reach unanimity.

The expected time to reach unanimity satisfies~\eqref{T_recursive_1}, which after identical steps of Section~\ref{sec:time_uncorr}, reduces to
\all{
0 &= \delta t 
 + \displaystyle \sum_k  \left( P_k^+ - P_k^- \right)     \rond{T(\vec{\rho})}{\rho_k} \delta \rho_k   
\nonumber \\
&+ 
\displaystyle \sum_k    \left( P_k^+  +    P_k^- \right)     \rondd{T(\vec{\rho})}{\rho_k} \fracc{\delta \rho_k^2}{2}  
.}{T_PDE_2_selfish}

Let us define two distinct sums:
\all{
\begin{cases}
\sigma_1 \deff \displaystyle \sum_k  \left( P_k^+ - P_k^- \right)     \rond{T(\vec{\rho})}{\rho_k} \delta \rho_k    \\ \\
\sigma_2 \deff \displaystyle \sum_k    \left( P_k^+  +    P_k^- \right)     \rondd{T(\vec{\rho})}{\rho_k} \fracc{\delta \rho_k^2}{2}  
.
\end{cases}
}{sigmas_def}

For the increment probability, we have to make adjustments to~\eqref{P_inc} as follows

\all{
P^+_k= n_k (1-\rho_k) \fracc{  k \displaystyle \sum_{\ell} \frac{\ell n_{\ell}}{\ave{k}} f(\ell) \rho_{\ell}}{f(k)+ k \displaystyle \sum_{\ell} \frac{\ell n_{\ell}}{\ave{k}} f(\ell) }.
}{P_inc_1_selfish}

Similarly, for the decrement probability we have
\all{
P^-_k= n_k  \rho_k  \fracc{ k \displaystyle \sum_{\ell} \frac{\ell n_{\ell}}{\ave{k}} f(\ell) (1-\rho_{\ell})}{f(k)+ k \displaystyle \sum_{\ell} \frac{\ell n_{\ell}}{\ave{k}} f(\ell) }.
}{P_dec_1_selfish}

From the increment and decrement probabilities,  we obtain
\all{
P^+_k - P^-_k = \fracc{k n_k }{\ave{k}} \fracc{\mu  -  \ave{kf(k)} \rho_k}{f(k)+  \fracc{\ave{kf(k)}}{\ave{k}} k} 
.}{P_differ_selfish}

Inserting this expression in $\sigma_1$ and also using the fact that~${\delta \rho_k =\frac{1}{N n_k}}$, we obtain
\all{
\sigma_1 = \displaystyle \sum_k  \fracc{k n_k }{N \ave{k}} \fracc{\mu  -  \ave{kf(k)} \rho_k}{f(k)+  \fracc{\ave{kf(k)}}{\ave{k}} k}  \rond{T(\vec{\rho})}{\rho_k}     
.}{sigma_1_temp_1}
By substituting $\mu$ by its explicit expression given in~\eqref{nu_def}, we can rewrite the numerator of the second factor of the summand as follows
\al{
\mu  -  \ave{kf(k)} \rho_k = \sum_{\ell} n_{\ell} \ell f(\ell) (\rho_{\ell}-\rho_k)
.}
Using this expression,~\eqref{sigma_1_temp_1} can be equivalently expressed as follows
\all{
\sigma_1 &= \displaystyle \sum_{k,\ell} \Bigg[  \fracc{k  }{N \ave{k}}  \rond{T(\vec{\rho})}{\rho_k}      \fracc{n_{\ell} \ell f(\ell) (\rho_{\ell}  - \rho_k)}{f(k)+  \fracc{\ave{kf(k)}}{\ave{k}} k} 
\Bigg]
.}{sigma_1_temp_2}

Using the chain rule, we have
\all{
\rond{T(\vec{\rho})}{\rho_k} &= \rond{T(\psi)}{\psi}
n_k  \bigg[  
 f^2(k)+ \fracc{\ave{kf(k)}}{\ave{k}} k f(k)\bigg] 
 \nonumber \\
 &= 
 \rond{T(\psi)}{\psi}
n_k  f(k) \bigg[  
 f(k)+ \fracc{\ave{kf(k)}}{\ave{k}} k  \bigg] 
.}{T_temp_selfish_1}
Now we plug this into~\eqref{sigma_1_temp_2}. Note that the numerator of the last factor  of the summand in~\eqref{sigma_1_temp_2} cancels out with the denominator of~\eqref{T_temp_selfish_1}, and we obtain
\al{
\sigma_1  =\frac{1}{N \ave{k}} \rond{T(\psi)}{\psi} \displaystyle \sum_{k,\ell}    k n_k f(k) \ell n_{\ell}  f(\ell)         (\rho_{\ell}  - \rho_k) 
.} 

Note that the summand is anti-symmetric with respect to the exchange of the indices $k$~and~$\ell$. So summing over all values of $k$~and~$\ell$ yields zero. So we have~${\sigma_1= 0}$. 

Returning to~\eqref{T_PDE_2_selfish}, we have the following simplified differential equation for the expected time to reach unanimity
\all{
0 = \delta t 
+ 
\displaystyle \sum_k    \left( P_k^+  +    P_k^- \right)     \rondd{T(\vec{\rho})}{\rho_k} \fracc{\delta \rho_k^2}{2}  
,}{T_PDE_3_selfish}

or equivalently, 
\all{
\displaystyle \sum_k    \left( P_k^+  +    P_k^- \right)     \rondd{T(\vec{\rho})}{\rho_k} \fracc{1}{2 n_k^2}=-N  
.}{T_PDE_4_selfish}

Combining~\eqref{P_inc_1_selfish} and~\eqref{P_dec_1_selfish}, we obtain

\all{
P^+_k+P^-_k= \frac{k n_k}{\ave{k}}   \fracc{ \mu+\ave{kf(k)} \rho_k - 2 \mu \rho_k 
}{f(k)+ k \frac{\ave{kf(k)}}{\ave{k}} }.
}{sum_of_Ps_1}

Employing the explicit expression for~$\mu$ as given in~\eqref{nu_def}, the numerator can be simplified as follows
\al{
 \mu+\ave{kf(k)} \rho_k - 2 \mu \rho_k  = \sum_{\ell} 
 n_{\ell} \ell f(\ell) ( \rho_{\ell} + \rho_k - 2 \rho_{\ell} \rho_k)
.}
So~\eqref{sum_of_Ps_1} becomes
\all{
P^+_k+P^-_k= \frac{k n_k}{\ave{k}}   \fracc{ 
\sum_{\ell} 
 n_{\ell} \ell f(\ell) ( \rho_{\ell} + \rho_k - 2 \rho_{\ell} \rho_k)
}{f(k)+ k  \frac{\ave{kf(k)}}{\ave{k}} }.
}{sum_of_Ps_2}
Plugging this into~\eqref{T_PDE_4_selfish}, we get

\all{
-N  =\displaystyle \sum_{k,\ell}      \rondd{T(\vec{\rho})}{\rho_k} \fracc{1}{2 n_k^2}  \frac{k n_k}{\ave{k}}   \fracc{ 
 n_{\ell} \ell f(\ell) ( \rho_{\ell} + \rho_k - 2 \rho_{\ell} \rho_k)
}{f(k)+ k \frac{\ave{kf(k)}}{\ave{k}} }  
.}{T_PDE_5_selfish}
%
%

To proceed, we confine ourselves to the case of linear clout, namely ${f(k)=k}$. In this case, from~\eqref{nu_def} we can see that $\mu$ and $\nu$ are the same, so $\mu$ is conserved. We can use the chain rule   to transfer the differentiation to $\mu$. Then~\eqref{T_PDE_5_selfish} transforms into

\all{
-N \ave{k} =\rondd{T(\mu)}{\mu}\displaystyle \sum_{k,\ell}     
 \fracc{1}{2  }  \frac{k^4 n_k}{1+\frac{\ave{k^2}}{\ave{k}}}    \left(\mu+\mu-2\frac{\mu^2}{\ave{k^2}}\right)  
.}{T_PDE_6_selfish}

Using the definition of~$\xi$ given in~\eqref{lambda_defs}, we can express~\eqref{T_PDE_6_selfish} as follows

\all{
   \rondd{T(\xi)}{\xi}  
\xi  \left( 1-   \xi  \right)  
 =- N \fracc{\ave{k^2}^2+\ave{k}\ave{k^2}}{\ave{k^4}}
.}{T_PDE_7_selfish}

Following the identical steps that led to~\eqref{T_FIN_full}, we obtain

\all{
  T(\xi)    
 = - N \frac{\ave{k^2}^2\!\! \!+ \!\! \ave{k} \!\! \ave{k^2}}{\ave{k^4}} \bigg[\xi \ln \xi   +(1-\xi)   \ln(1-\xi)  \bigg] 
.}{T_FIN_full_selfish}

This is longer than the consensus time given by~\eqref{T_FIN_full}, due to  the extra term in the numerator that stems from the confidence of nodes.

\begin{figure}
  \centering
  \includegraphics[width=1\columnwidth]{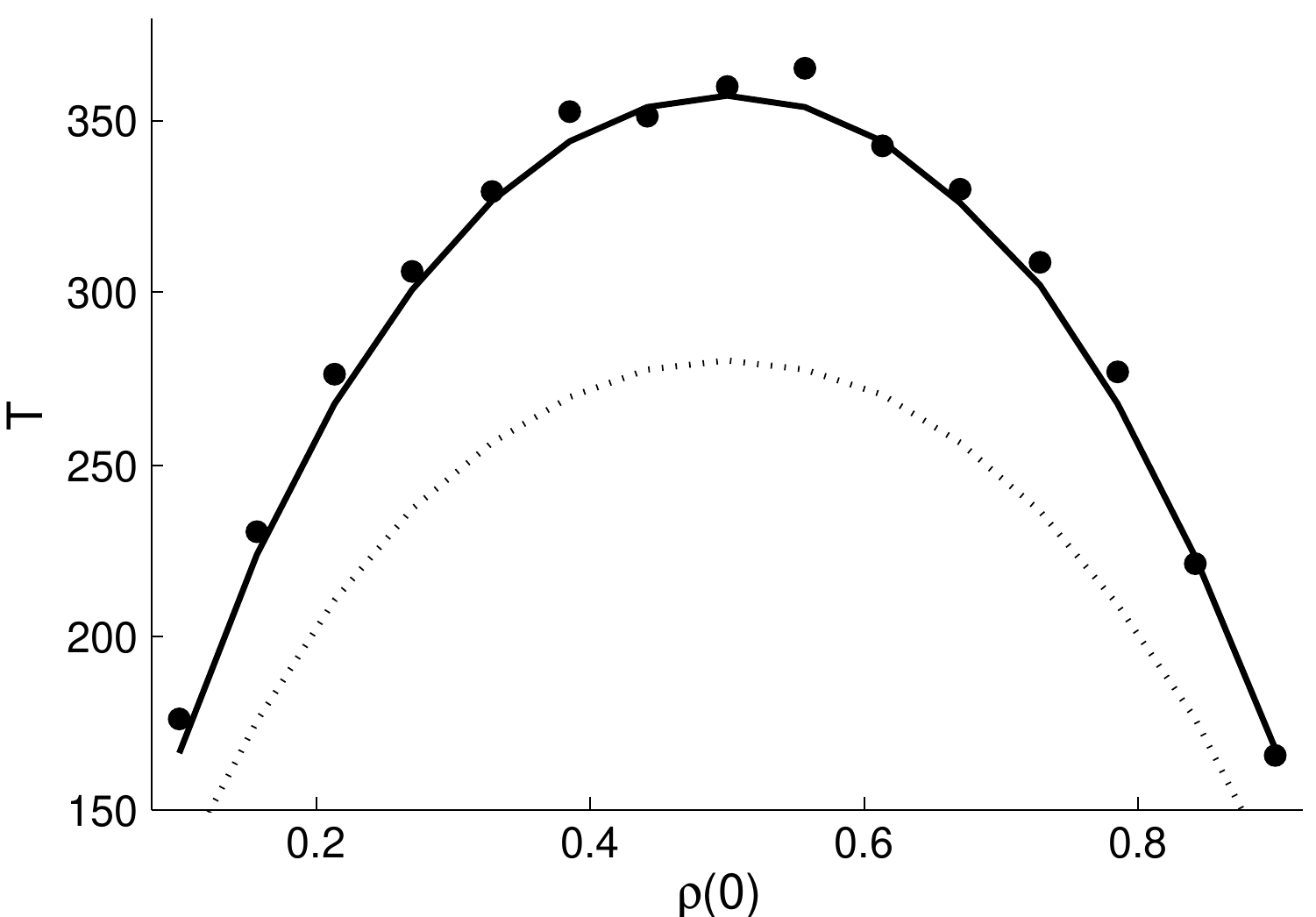}
  \caption[Figure ]%
   {  Expected time to reach unanimity  as a function of initial density of nodes with state $+1$ for the case of linear clout~${f(k)=k}$. The markers are simulation results and the solid line is theoretical prediction given by~\eqref{T_FIN_full_selfish}. The dashed line shows the consensus time when nodes have no confidence. It can be observed that confidence slows down convergence. The underlying graph has power law degree distribution ${n_k \sim k^{-3}}$ as proposed in~\cite{BA} with $m=20$. The network has 1500 nodes and the results are averaged over 100 Monte Carlo trials.}
\label{confident_T}
\end{figure}

Theoretical prediction and simulation results are presented in Fig.~\ref{confident_T}. The expected consensus time for the absence of confidence is also depicted, and it can be seen that confidence slows down the convergence of the system towards unanimity, as predicted by~\eqref{T_FIN_full_selfish}.

For the expected time to consensus conditional on the final state, similar steps that led to~\eqref{T_COND_FIN} apply here, and we arrive at

\all{
\begin{cases}
T^u(\xi)= - N \frac{\ave{k^2}^2 +   \ave{k}   \ave{k^2}}{\ave{k^4}} \fracc{1-\xi}{\xi} \ln(1-\xi) \\ \\
T^d(\xi)=- N \frac{\ave{k^2}^2 +  \ave{k}   \ave{k^2}}{\ave{k^4}}  \fracc{\xi}{1-\xi} \ln(\xi) .
\end{cases} 
}{T_COND_FIN_selfish}

\begin{figure}
  \centering
     \includegraphics[width=1\columnwidth]{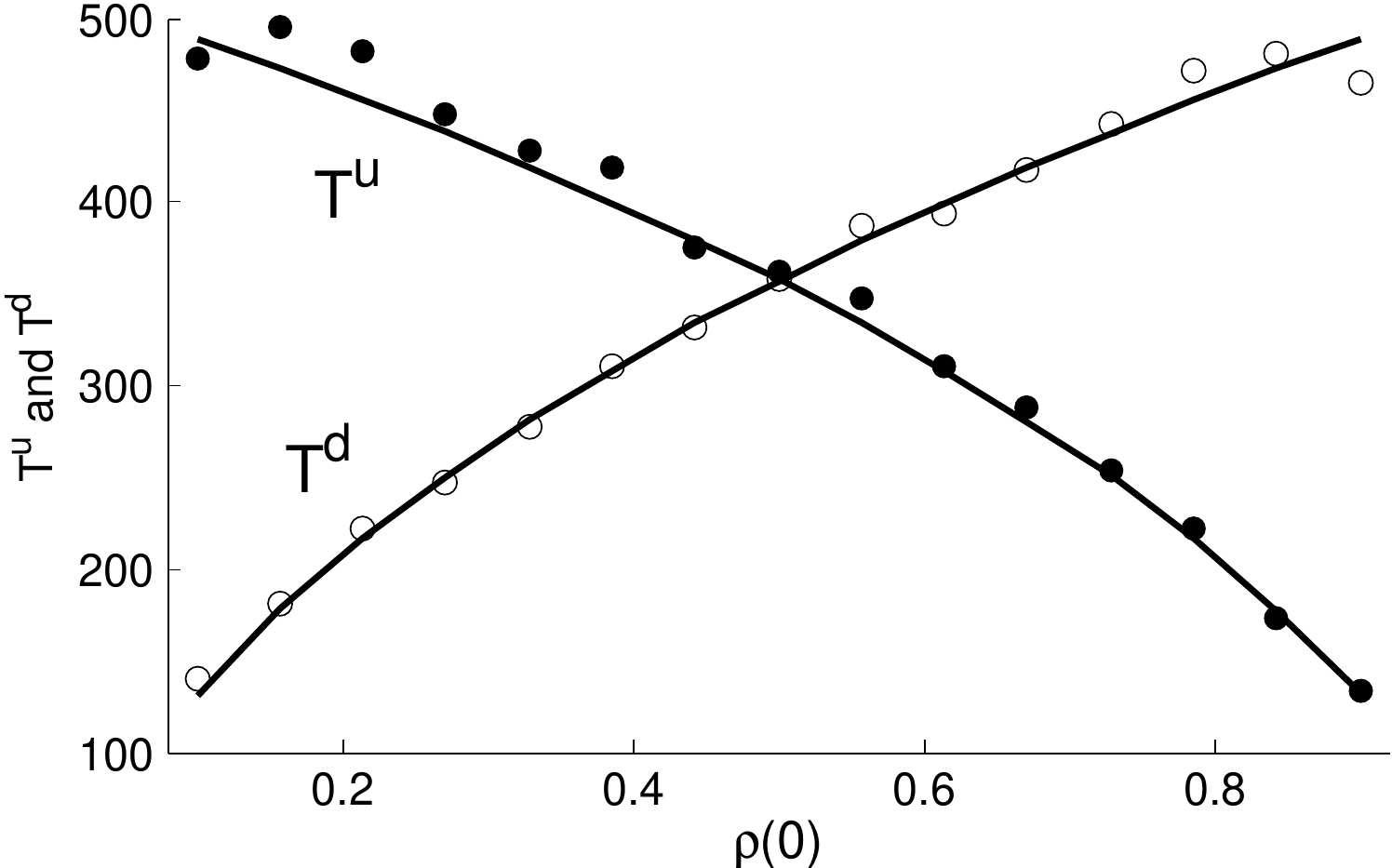}
  \caption[Figure ]%
   { Expected time to reach unanimity conditional on states, for the case of linear clout~${f(k)=k}$. Conditional consensus times are depicted as a function of initial density of nodes with state $+1$. The markers are simulation results and the solid lines are theoretical predictions given by~\eqref{T_COND_FIN_selfish}. The underlying graph has power law degree distribution ${n_k \sim k^{-3}}$ as proposed in~\cite{BA} with $m=20$. The network has 1500 nodes and the results are averaged over 100 Monte Carlo trials.    }
\label{confident_TU}
\end{figure}

Theoretical predictions for the conditional consensus times are presented along with simulation results in Fig.~\ref{confident_TU}.

\section{Irreversible Dynamics}

As discussed in the introduction, the voter model is not applicable to marketing problems where once a node sees a film (correspondingly, adopts $+1$), it cannot go back, i.e., it cannot un-see it. Here we consider an irreversible version of the voter model. 
%
The probability of an increment in the population of adopters of state $+1$ with degree $k$ is 
\al{
P^+_k= n_k (1-\rho_k) \fracc{\displaystyle \sum_{\ell} \frac{\ell n_{\ell}}{\ave{k}} f(\ell) \rho_{\ell}}{\displaystyle \sum_{\ell} \frac{\ell n_{\ell}}{\ave{k}} f(\ell) }
= n_k (1-\rho_k) \fracc{\mu}{\ave{kf(k)}}
.
}
So for the evolution of densities we have
\all{
\frac{d \rho_k}{dt} = (1-\rho_k) \fracc{\mu}{\ave{kf(k)}}
.}{ro_dot_irrev}
Taking the time derivative of $\mu$ and using~\eqref{ro_dot_irrev}, we obtain
\al{
\dot{\mu}= \DS \sum_k n_k k f(k) \dot{\rho_k}
= \fracc{\mu}{\ave{kf(k)}}   \DS \sum_k n_k k f(k) (1-\rho_k)
.}
This can be simplified to yield the following differential equation 
\al{
\dot{\mu}=
\fracc{\mu}{\ave{kf(k)}} \bigg[ \ave{kf(k)}-\mu \bigg] 
.}
This can be equivalently expressed as follows
\al{
d \mu \left[ \fracc{1}{\mu}+\fracc{1}{\ave{kf(k)}-\mu} \right]= dt 
,}
which can be integrated to give
\al{
\ln\left[  \fracc{\mu(t)}{\mu(0)}
 \fracc{\ave{kf(k)}-\mu(0)}{\ave{kf(k)}-\mu(t)}  \right] =t
.}
Rearranging the terms, we can find $\mu(t)$ as a function of time
\all{
\mu(t)= \fracc{\ave{kf(k)} \mu(0)}{\ave{kf(k)} e^{-t} + \mu(0)(1-e^{-t})}
.}{mu_t_irrev}

\begin{figure}
  \centering
  \includegraphics[width=1\columnwidth]{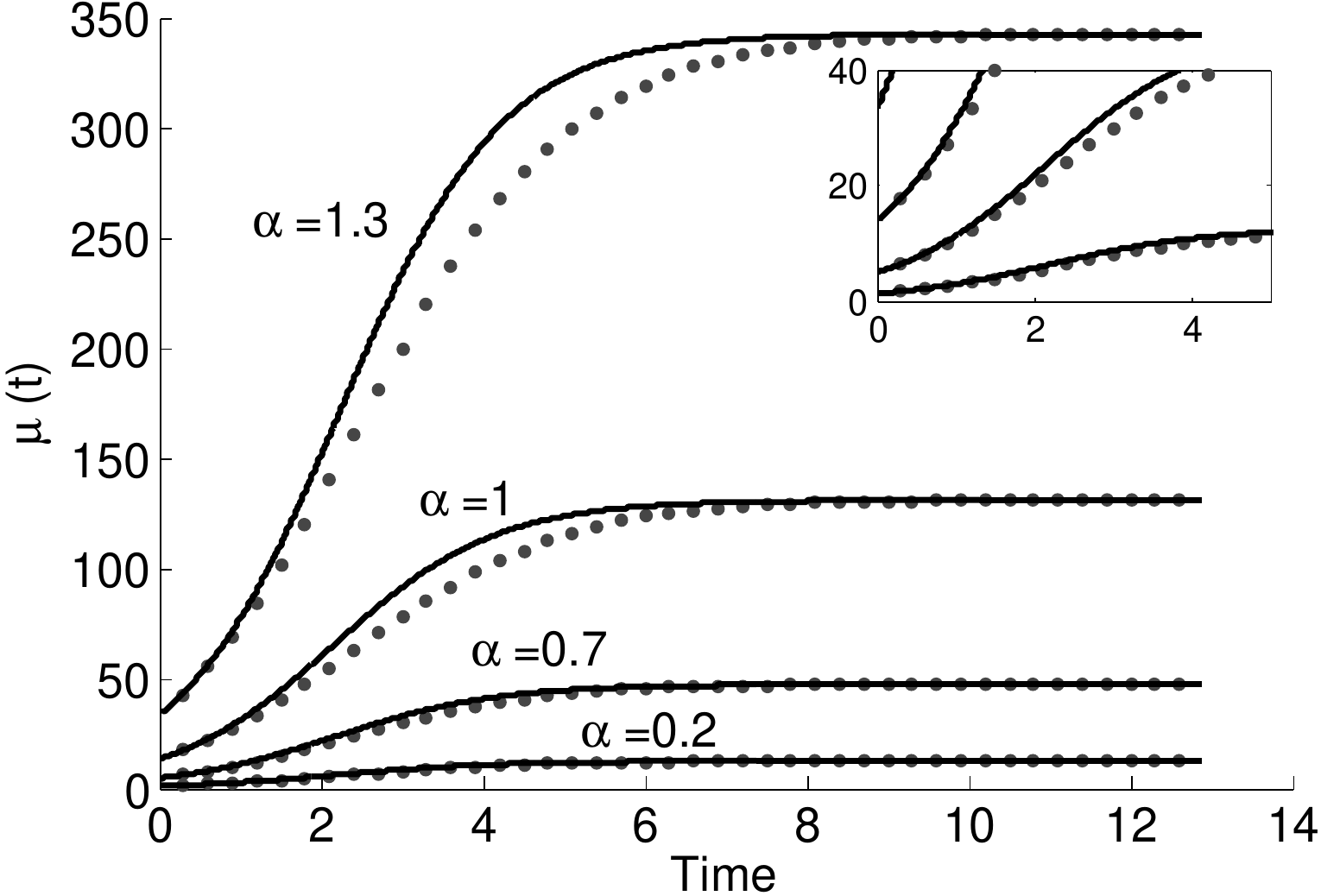}
  \caption[Figure ]%
   {  Time evolution of $\mu(t)$. The markers are simulation results and the solid lines are theoretical predictions given by~\eqref{mu_t_irrev}. The underlying graph has power law degree distribution ${n_k \sim k^{-3}}$ as proposed in~\cite{BA} with $m=20$. The special case of ${f(k)=k^{\alpha}}$ is considered, for ${\alpha=0.2,0.7, 1, 1.3}$. The network has 1500 nodes and the results are averaged over 100 Monte Carlo trials. The inset provides a more visible depiction for the early stage of the dynamics for the two smallest $\alpha$s.}
\label{fig_mu_t}
 \vspace{-4mm}
\end{figure}

Theoretical prediction and simulation results for the evolution of $\mu(t)$ are presented in Fig.~\ref{fig_mu_t} for the case of ${f(k)=k^{\alpha}}$ for ${\alpha=0.2,0.7, 1, 1.3}$. As can be seen from the figure, higher values of $\alpha$ converge faster towards $+1$. This signifies the role of high-degree nodes in spreading the influence throughout the system.

Getting back to~\eqref{ro_dot_irrev} and inserting the expression for $\mu(t)$ obtained in~\eqref{mu_t_irrev}, we have the following differential equation for the evolution of densities:
\all{
\frac{d \rho_k}{dt} = (1-\rho_k) \fracc{  \mu(0)}{\ave{kf(k)} e^{-t} + \mu(0)(1-e^{-t})}
.}{ro_dot_irrev_2}
To solve this equation, let us temporarily use the following definition for brevity
\al{
g(t) \deff \fracc{  \mu(0)}{\ave{kf(k)} e^{-t} + \mu(0)(1-e^{-t})}
.}
Then we can rewrite~\eqref{ro_dot_irrev_2} as follows
\al{
\dot{\rho_k} + g(t) \rho_k = g(t)
.}
This is a linear first order differential equation. Multiplying both sides by the integration factor $\phi(t)$ in order to make both sides equal to $\dot{\rho_k \phi(t)}$, we find that ${\phi(t)= \exp \int g(t) dt}$. Then the complete solution is 
\all{
\rho_k(t)= \frac{1}{\exp \int g(t) dt } \left[\DS \int g(t) e^{\int g(t) dt} dt +C \right] 
,}{recipe_1}
where $C$ is a constant determined from the initial conditions. Note that integrations are indefinite.

We have to compute two integrals, namely ${\int g dt}$ and ${\int g e^g dt}$. For the first integral we have
\all{
\DS \int g(t) dt &= \DS \int       \fracc{  \mu(0)}{\ave{kf(k)} e^{-t} + \mu(0)(1-e^{-t})}   dt
\nonumber \\
&= \ln \left[  \mu(0) e^t + \ave{k f(k)}-\mu(0) \right] 
 =  \ln \left[  \frac{\mu(0)}{g(t)}  e^t \right] 
.}{g_int}
Then we perform the second integration as follows
\all{
\DS \int g(t) e^{\int g(t) dt} dt  = 
\DS \int \mu(0) e^t dt = \mu(0) e^t
.}{g_int_2}
Inserting the results of~\eqref{g_int} and~\eqref{g_int_2} into~\eqref{recipe_1}, we get
\al{
\rho_k (t)= \fracc{  \mu(0)+C e^{-t}}{\ave{kf(k)} e^{-t} + \mu(0)(1-e^{-t})} 
.}
The constant $C$ is obtained by imposing the initial condition at~${t=0}$.     The final result is
\all{
\rho_k (t) =   \fracc{   \ave{kf(k)} \rho_k(0)   e^{-t}+\mu(0) (1-e^{-t})}{\ave{kf(k)} e^{-t} + \mu(0)(1-e^{-t})} 
.}{rho_k_final_irrev}

This lead us to the total proportion of nodes of state $+1$, which we denote by $\rho$. Multiplying this equation by $n_k$ and summing over all $k$, we obtain
\all{
\rho (t) =    \fracc{   \ave{kf(k)} \rho(0)   e^{-t}+\mu(0) (1-e^{-t})}{\ave{kf(k)} e^{-t} + \mu(0)(1-e^{-t})} 
.}{rho_tot_irrev}

\subsection{Time to Unanimity}
Now let us find the expected time it takes for the system to reach a complete unanimous state with every node adopting $+1$. Denote this time by $T$. Note that, since we have taken the time unit in a way that on average at each timestep every node is selected once. Denote by $T_{N-1}$ the expected time at which ${N-1}$ nodes have state $+1$. Then the following recursion holds:
\all{
T=T_{N-1}+1
.}{T_recur_irrev}
To obtain $T_{N-1}$, we use~\eqref{rho_tot_irrev} for ${\rho=1-\frac{1}{N}}$ as follows
\al{
1-\frac{1}{N}=    \fracc{   \ave{kf(k)} \rho(0)   e^{-t}+\mu(0) (1-e^{-t})}{\ave{kf(k)} e^{-t} + \mu(0)(1-e^{-t})} 
.}
This can be rearranged to give
\al{
\exp (-T_{N-1}) = \fracc{\mu(0)}{N\big[1-\rho(0)\big]\ave{kf(k)} + \mu(0)-\ave{kf(k)}}
.}
Taking the logarithm and plugging the result in~\eqref{T_recur_irrev} yields
\all{
T= 1+ \ln \left[ 1+ \fracc{\ave{kf(k)}}{\mu(0)} \big[N-N\rho(0)-1 \big]  \right] 
.}{T_FIN_irrev}

\begin{figure}
  \centering
  \includegraphics[width=1\columnwidth]{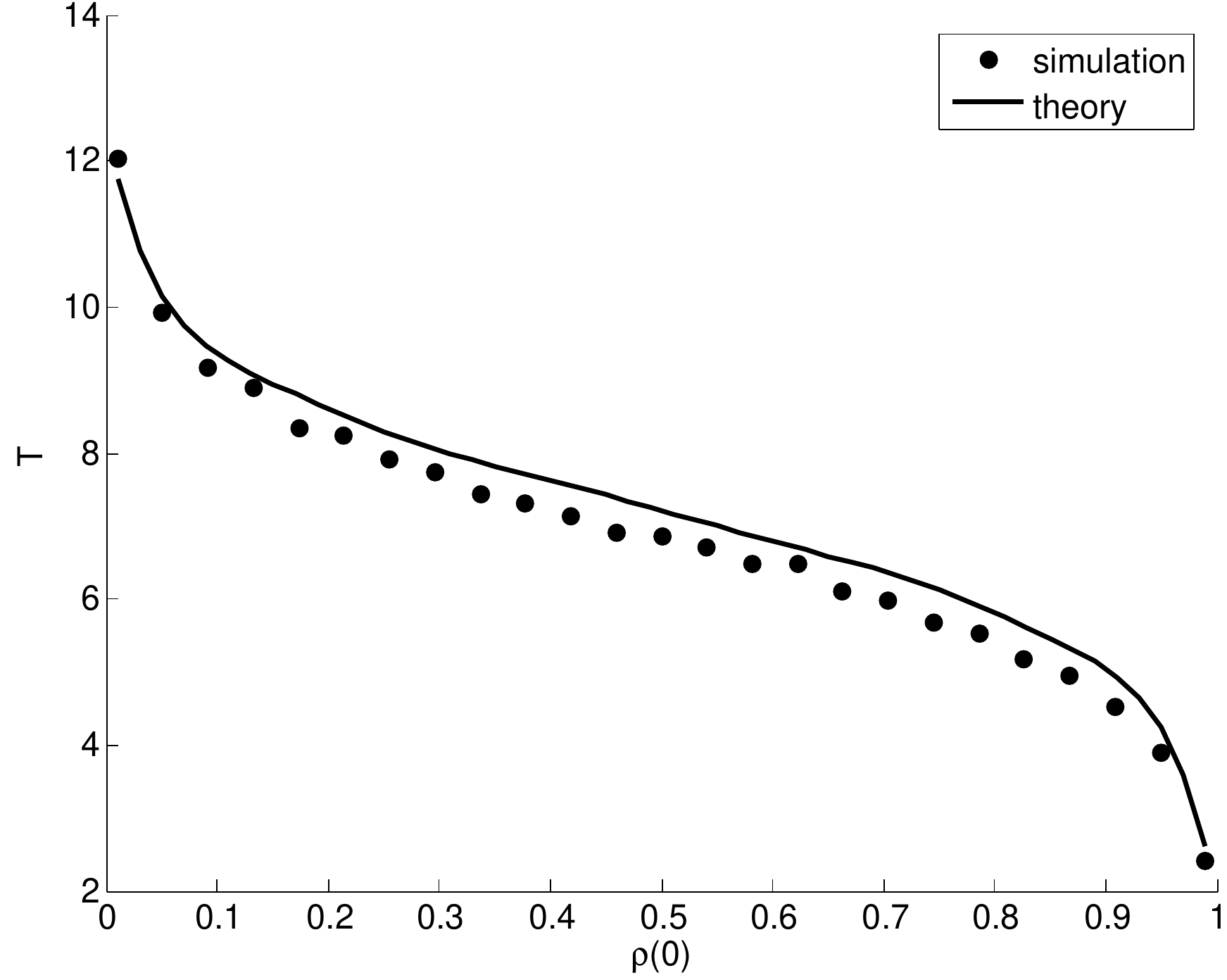}
  \caption[Figure ]%
   {  Time to reach unanimity for the irreversible model, as a function of the initial density of adopters. Theoretical prediction is given by~\eqref{T_FIN_irrev}. The case of $f(k)=k^{0.8}$ is considered for simulation purposes.    The underlying graph has power law degree distribution ${n_k \sim k^{-3}}$ as proposed in~\cite{BA}, with $m=20$ and 500 nodes. The results are averaged over 100 Monte Carlo trials. 
   }
\label{T_rho_irrev_fig}
\end{figure}

Theoretical prediction and simulation results are illustrated in Fig.~\ref{T_rho_irrev_fig}.

With the help of~\eqref{T_FIN_irrev} we can also estimate the expected time for the system to reach unanimity in the case of a single initial mutant. Denote the degree of the mutant by $z$.  Considering a network with power-law degree distribution and the special case of linear clout, namely ${f(k)=k}$, we have~${\mu(0)= \frac{z^2}{N} }$. 
 For scale-free graphs ${\ave{kf(k)}}$ grows at most polynomially in $N$~\cite{SOOD}, so  we have
\all{
T \sim \ln N
.}{T_irrev_sim}

\begin{figure}
  \centering
  \includegraphics[width=1\columnwidth]{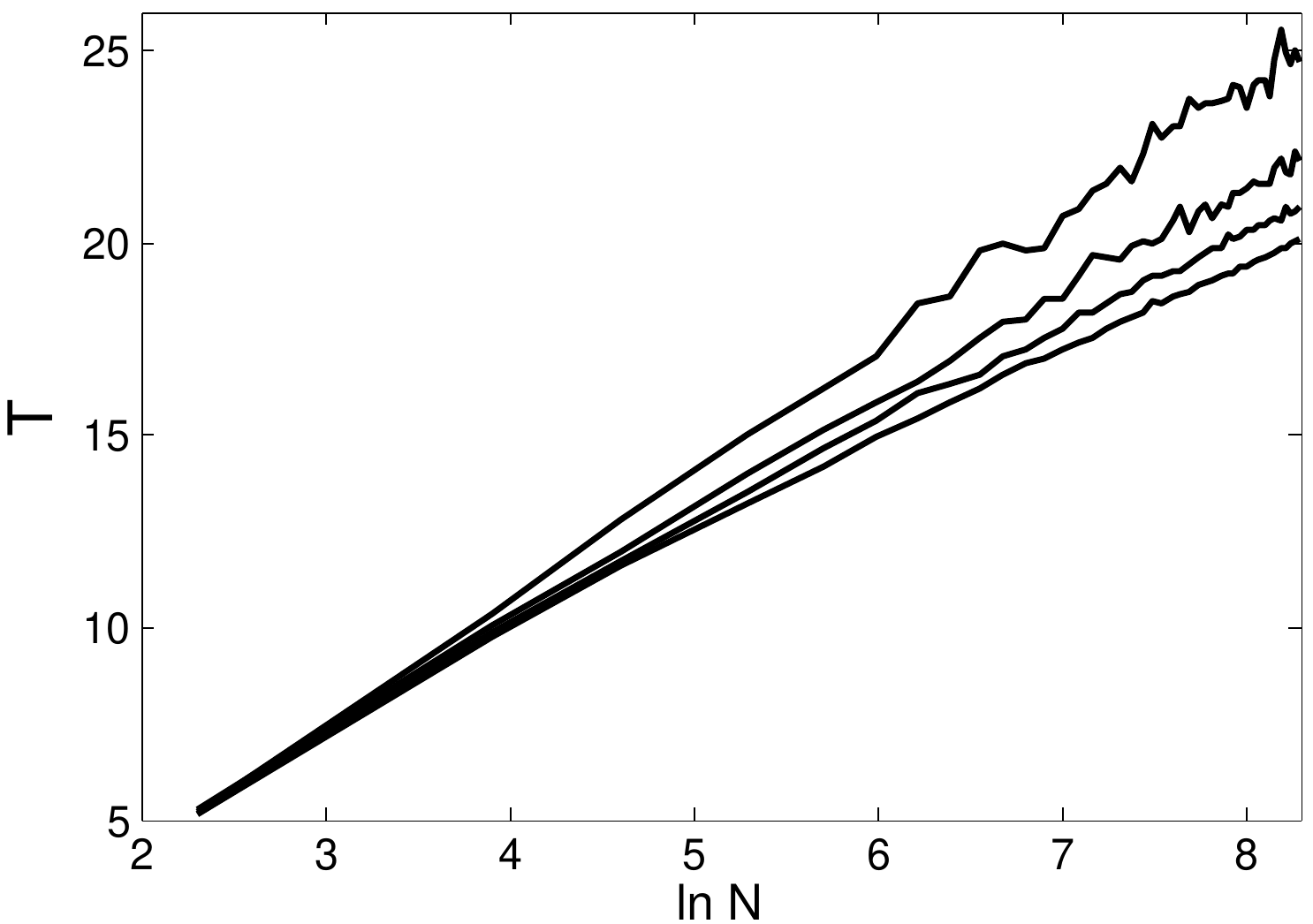}
  \caption[Figure ]%
   {  Consensus time as a function of logarithm of network size for the case of ${f(k)=k^{\alpha}}$ with ${\alpha=0.2,0.7, 1,1.3}$ (from top to bottom, i.e., the top-most curve pertains to ${\alpha=0.2}$). As~\eqref{T_irrev_sim} predicts, the curves are linear. The results are averaged over 1000 Monte Carlo simulations.
   }
\label{fig_irrev_mutant}
 \vspace{-4mm}
\end{figure}

Simulation results for the behavior of consensus time is depicted in Fig.~\ref{fig_irrev_mutant}. Consensus time is plotted against logarithm of network size, for the case of ${f(k)=k^{\alpha}}$ with ${\alpha=0.2,0.7, 1,1.3}$. All curves are visibly linear, in agreement with~\eqref{T_irrev_sim}.

\section{Summary}

The voter model is a simple stochastic process frequently used to emulate   opinion dynamics   on social networks. One of its oversimplifications is homogeneity of influence, i.e., each node is influenced equally by all of its neighbors, regardless of their characteristics. Another is that nodes, even those with large degrees, have no confidence; their decisions are solely based upon the actions of their neighbors. The third drawback of the voter model we alluded to in this paper is more pragmatic. It pertains to marketing applications. Once a node adopts the state that corresponds to, for example,  seeing a film or adopting a technology, they cannot go back. 

The focus of this paper was to address these issues. We endowed nodes with  status, as a function of their degrees. Confidence was incorporated into the model by making nodes treat their own selves as a neighbor, so that the higher the degree of a node is, the more stubborn it gets against altering its state. We also studied another extension to the voter model, where nodes can flip from state $-1$ to $+1$ but not the other way around. 

In each case, we studied the dynamics of the system and compared our theoretical predictions with simulation results.

%
%
%
%
%
%

 \begin{spacing}{0.88}
\bibliographystyle{IEEEtran}
\bibliography{myref}
\end{spacing}

\end{document}